\documentstyle[12pt]{article}
\def\dir{}     
\def\figcond{1}     
\def\figsty{0}      

\newcommand{\msection}[1]{%
\setcounter{equation}{0}
\section[]{#1}}

\renewcommand{\today}{\ifcase\month\or
 Jan.\or Feb.\or Mar.\or Apr.\or May\or Jun.\or
 Jul.\or Aug.\or Sep.\or Oct.\or Nov.\or Dec.\fi
 \space\number\day, \number\year}

\def\authname{M.Harada, Y.Kikukawa, T.Kugo and H.Nakano}
\newcount\figsubcountss \figsubcountss=0
\ifnum\figcond>0
\input epsf
  \ifnum\figsty>0
  \fi
\fi
\def\figinsert#1#2#3{
\ifnum\figcond>0
  \ifnum\figsty>0
    \ifnum\figsubcountss=0
      \immediate\write9{\noexpand\catcode`\noexpand\@=11}
      \immediate\write9{\noexpand\newpage}
      \immediate\write9{\noexpand\pagestyle {empty}}
      \immediate\write9{\noexpand\section* {Figure Captions}}
      \immediate\write9{\noexpand\begin {enumerate}}
      \immediate\write9{\noexpand\renewcommand {\noexpand\theenumi}{}}
      \immediate\write9{\noexpand\begin {enumerate}}
      \immediate\write9{\noexpand\renewcommand
        {\noexpand\theenumii}{\noexpand\arabic {enumii}}}
      \immediate\write9{\noexpand\renewcommand {\noexpand\labelenumii}
        {Fig. \noexpand\arabic {enumii}:}}
      \immediate\write10{\noexpand\newpage}
      \ifnum\figsty>1
        \immediate\write10{\noexpand\pagestyle {headings}}
        \immediate\write10{\noexpand\setcounter {page}{1}}
        \immediate\write10{\noexpand\renewcommand {\noexpand\thepage}
            {Figure \noexpand\arabic{page} -- \noexpand\authname}}
      \else
        \immediate\write10{\noexpand\pagestyle {empty}}
      \fi
    \fi
    \global\advance\figsubcountss by 1
    \immediate\write10{\noexpand\begin {figure}[p]}
    \immediate\write10{\noexpand\begin {center}}
    \immediate\write10{\noexpand\ }
    \immediate\write10{\noexpand\epsfbox {\dir #1}}
    \ifnum\figsubcountss=3
      \immediate\write10{\noexpand\setlength %
        {\noexpand\unitlength}{0.1pt}}
      \immediate\write10{\noexpand\begin {picture}(994,335)(0,-1000)}
      \immediate\write10{\noexpand\put (1000,-450) %
        {\noexpand\makebox (0,0)[lb]{\noexpand\large$\noexpand\infty$}}}
      \immediate\write10{\noexpand\end {picture}}
    \fi
    \immediate\write10{\noexpand\ \noexpand\\}
    \ifnum\figsty<2
      \immediate\write10{\noexpand\vspace {1cm}}
      \immediate\write10{Figure \noexpand\ref {#3}}
    \fi
    \immediate\write10{\noexpand\end {center}}
    \immediate\write10{\noexpand\end {figure}}
    \immediate\write10{\noexpand\newpage}
   \let\save=\ref \let\ref=0 \let\savec=\cite \let\cite=0
    \immediate\write9{\noexpand\item #2}
   \let\ref=\save \let\cite=\savec
    \immediate\write9{\noexpand\label {#3}}
    \begin {figure}[htbp]
    \begin{center}
    \fbox{Fig. \ref{#3}}
    \end{center}
    \end {figure}
  \else
    \global\advance\figsubcountss by 1
    \begin{figure}[htbp]
    \begin{center}
    \ \epsfbox{\dir #1}
    \ifnum\figsubcountss=3
      \setlength{\unitlength}{0.1pt}
      \begin{picture}(994,335)(0,-1000)
      \put(1000,-450){\makebox(0,0)[lb]{\large$\infty$}}
      \end{picture}
      \vspace{-30pt}
    \fi
    \vspace{20pt}
    \caption []{#2 \label {#3}}
    \end{center}
    \end{figure}
  \fi
\else
  \begin {figure}[htbp]
  \begin{center}
  \fbox{Fig. \ref{#3}}
  \caption []{#2 \label {#3}}
  \end{center}
  \end {figure}
\fi
}
\def\Closeout#1{%
   \immediate\closeout#1}
\def\figepsfout{
\Closeout10
  \immediate\write9{\noexpand\end {enumerate}}
  \immediate\write9{\noexpand\end {enumerate}}
  \immediate\write9{\noexpand\catcode`\noexpand\@=12}
\Closeout9
\input \jobname.cap
\input \jobname.fis}

\catcode`\@=11
\newbox\tempboxa
\newdimen\captionboxsubcount
\def\capsize#1{\captionboxsubcount=#1pt}
\newdimen\captionboxsub
\captionboxsub=\hsize \advance\captionboxsub by -\captionboxsubcount
\advance\captionboxsub by -\captionboxsubcount
\long\def\@makecaption#1#2{
 \setbox\@tempboxa\hbox{#1: #2}
 \ifdim \wd\@tempboxa >\captionboxsub
\rightskip=\captionboxsubcount \leftskip=\captionboxsubcount #1: #2
\else \hbox to\hsize{\hfil\box\@tempboxa\hfil}
 \fi}
\catcode`\@=12
\capsize{30}

\newcommand{\abs}[1]{\left\vert{#1}\right\vert}
\newcommand{\nonum}{\nonumber\\}
\newcommand{\be}{\begin{equation}}
\newcommand{\ee}{\end{equation}}
\newcommand{\ba}{\begin{eqnarray}}
\newcommand{\ea}{\end{eqnarray}}
\newcommand{\bd}{\begin{displaymath}}
\newcommand{\ed}{\end{displaymath}}
\newcommand{\Nc}{N_{\rm C}}
\newcommand{\Nf}{N_{\rm f}}

\newcommand{\p}{\partial}	

\newcommand{\nf}{n_{\rm f}}
\newcommand{\alz}{\alpha_0}
\newcommand{\alc}{\alpha_{\rm c}}

\newcommand{\hz}{h_0}
\newcommand{\kz}{k_0}
\newcommand{\tcut}{t_\Lambda}

\newcommand{\mmd}{m_{\rm {\scriptscriptstyle MD}}}
\newcommand{\ycomp}{y_{\Lambda}}

\newcommand{\wbycomp}{\wb{y}_{\Lambda}}
\newcommand{\wblamcomp}{\wb{\lambda}_{\Lambda}}

\def\MS{\overline{\rm MS}}

\def\gHY{{gauge-Higgs-Yukawa}}
\def\ie{{\it i.e.}}
\def\eg{{\it e.g.}}

\def\aoac{\alpha/\alc}
\def\cob{c/b}

\def\runn{\eta(t)}
\def\run#1{\eta^{#1}(t)}
\newcommand{\runs}{\eta}

\newcommand{\wb}{\overline} 
\newcommand{\raw}{\rightarrow}
\newcommand{\lra}{\ \longrightarrow\ }

\def\MF{M_{{\rm F}}}
\def\wbMF{\wb{M}_{{\rm F}}}

\newcommand{\sD}{\rlap / \!D}
\newcommand{\BKMN}{\eta}

\newcommand{\tc}{t_{{\rm {\scriptscriptstyle C}}}}
\newcommand{\muc}{\mu_{{\rm {\scriptscriptstyle C}}}}

\catcode`\@=11 
\def\lsim{\mathrel{\mathpalette\@versim<}}
\def\gsim{\mathrel{\mathpalette\@versim>}}
\def\@versim#1#2{\vcenter{\offinterlineskip
        \ialign{$\m@th#1\hfil##\hfil$\crcr#2\crcr\sim\crcr } }}
\catcode`\@=12 

\newcommand{\LQCD}{\Lambda_{\rm QCD}}

\def\NP#1{{\it Nucl}.~{\it Phys}. {{\bf #1}},}
\def\PL#1{{\it Phys}.~{\it Lett}. {{\bf #1}},}
\def\PR#1{{\it Phys}.~{\it Rev}. {{\bf #1}},}

\def\PRep#1{{\it Phys}.~{\it Report} {{\bf #1}},}
\def\PTP#1{{\it Prog}.~{\it Theor}.~{\it Phys}. {{\bf #1}},}
\def\MPL#1{{\it Mod}.~{\it Phys}.~{\it Lett}. {{\bf #1}},}
\def\IJMP#1{{\it Int}.~{J}.~{\it Mod}.~{\it Phys}. {{\bf #1}},}
\def\RMP#1{{\it Rev}.~{\it Mod}.~{\it Phys}. {{\bf #1}},}

\begin{document}

\thispagestyle{empty}
\begin{titlepage}

\begin{flushright}
\begin{minipage}[t]{4cm}
\begin{flushleft}
KUNS-1278 \\
HE(TH) 94/10\\
NIIG-DP-94-2\\
hep-ph/9407398 \\
July, 1994
\end{flushleft}
\end{minipage}
\end{flushright}

\vspace{0.7cm}

\begin{center}
\Large\bf
Nontriviality of Gauge-Higgs-Yukawa System
and Renormalizability of Gauged NJL Model
\end{center}

\vfill

\begin{center}
{\large
Masayasu {\sc Harada}\footnote{
Fellow of the Japan Society for the
Promotion of Science for Japanese Junior Scientists.\\\indent%
\ $\,$address after August 20: Dept.\ of Phys.\ %
Syracuse University, Syracuse, New York 13244-1130
},
Yoshio {\sc Kikukawa},
Taichiro {\sc Kugo}
}
\\
{\it Department of Physics, Kyoto University, Kyoto 606-01, Japan}
\\
and
\\
{\large
Hiroaki {\sc Nakano}\footnote{e-mail address: {\tt nakano@niigt.kek.jp}}
}
\\
{\it Department of Physics, Niigata University, Niigata 950-21, Japan}
\end{center}

\vfill

\vskip 1cm

\begin{abstract}
In the leading order of a modified $1/\Nc$ expansion,
we show that a class of gauge-Higgs-Yukawa systems
in four dimensions give non-trivial
and well-defined theories in the continuum limit.
The renormalized Yukawa coupling $y$
and the quartic scalar coupling $\lambda$ have to
lie on a certain line in the $(y,\lambda)$ plane
and the line terminates at an upper bound.
The gauged Nambu--Jona-Lasinio (NJL) model
in the limit of its ultraviolet cutoff going to infinity,
is shown to become equivalent to the gauge-Higgs-Yukawa system
with the coupling constants just on that terminating point.
This proves the renormalizability of
the gauged NJL model in four dimensions.
The effective potential for the gauged NJL model is calculated
by using renormalization group technique
and confirmed to be consistent with the previous result
by Kondo, Tanabashi and Yamawaki
obtained by the ladder Schwinger-Dyson equation.
\end{abstract}
\end{titlepage}

\setcounter{footnote}{0}
\msection{Introduction}

It is generally a difficult problem
whether a theory defined with an ultraviolet cutoff
(such as in lattice formulation) really has a well-defined
(\ie, finite) continuum limit which is not a free theory.
If the theory becomes necessarily free in the continuum limit,
the theory is called trivial and
if it gives an interacting theory, it is called nontrivial.
This {\it triviality} or {\it nontriviality} is always
a problem independently of whether the theory is
perturbatively renormalizable or not.
But, if a theory is perturbatively non-renormalizable and
nevertheless gives a nontrivial continuum limit,
people prefer to call it (non-perturbatively) {\it renormalizable}
rather than simply calling it nontrivial.
We also follow this terminology in this paper.

We discuss two problems in this paper.
One is the triviality problem of a class of
gauge-Higgs-Yukawa systems in four dimensions
which are of course (perturbatively) renormalizable.
We examine and clarify when they can give well-defined,
nontrivial theories in the continuum limit
(\ie, when the ultraviolet cutoff goes to infinity).
Another is the renormalizability of
gauged Nambu--Jona-Lasinio (NJL) models in four dimensions
which are (perturbatively) non-renormalizable.
We note that the gauged NJL models are equivalent
to special cases of \gHY\ systems at the stage
with the ultraviolet cutoff kept finite.
Considering the limit of the cutoff going to infinity,
we can show that, under a certain condition,
the gauged NJL models give well-defined continuum limits
which are equivalent to specific nontrivial \gHY\ theories.
Namely, gauged NJL models become renormalizable
in the sense of above terminology.

Kondo, Tanabashi and Yamawaki\cite{KTY:PTP} (KTY)
have studied the gauged NJL
model for the fixed (non-running) gauge coupling case and
have shown the renormalizability within
the ladder approximation of the Schwinger-Dyson (SD) equation.
This
conclusion has further been supported by their recent work with
Shibata\cite{KSTY} in which they analyzed the flow of the renormalized
Yukawa coupling and mass parameter in the corresponding \gHY\
theory.
These works are, however, restricted to the fixed gauge coupling case,
and
moreover their arguments are based on a particular technique of
ladder SD equation whose nature of approximation is rather unclear.
It is therefore necessary to do clearer analysis based on a more
sound technique. We adopt the renormalization group equation
(RGE) as our basic tool in this paper
and directly examine them in a definite approximation scheme.

We spell out here about an assumption we take.
{}For the present purpose,
we should ideally analyze the RGE's
in a Wilson's sense\cite{Wilson-Polchinski}
which trace the RG flow of the coupling constants
of the cutoff theories.
However, we actually do not work directly in the cutoff theories.
Instead, we analyze the RGE's in the continuum theories,
but assume that the coupling constants renormalized at $\mu $ there
can be identified with the bare coupling constants
in the cutoff theory with cutoff $\Lambda =\mu $,
at least when the renormalization point $\mu $ becomes very large.
The validity of this assumption, however, would generally
depend on how the renormalized coupling constants are defined
in the continuum theory.
The authors of ref.~\cite{BKMN:CriticalInst}
have explicitly worked out on this point
in a Higgs-Yukawa system and seen the following:
such identification is valid for coupling constants
corresponding to operators of dimension four.
But a care had to be taken for the mass parameter
to which quadratic divergences are relevant.
We should use the mass parameter
in the mass-dependent renormalization scheme
in order to identify it with that in the cutoff theory.
Our assumption here is that the situation is the same
also in the present gauged Higgs-Yukawa systems.

To analyze the above stated problems,
we work in the leading order in $1/\Nc$ expansion
combined with the perturbation with respect to a small,
asymptotically free gauge coupling constant $g^2$.
We also use an additional assumption that
the pure $\lambda\phi^4$ theory is trivial\cite{Callaway:PRep}.
Then we show that the followings should hold in order
for the gauge-Higgs-Yukawa system to give a nontrivial
and well-defined theory.
(i)~{\it
There exists a functional relation between
the renormalized Yukawa coupling $y(\mu)$ and
quartic scalar coupling $\lambda(\mu)$.
}
Namely, these couplings should lie on a certain line
in the $(y,\lambda)$ plane.
(ii)~{\it
There exists a nonzero, finite upper bound
for the renormalized Yukawa coupling.
}
The line in the $(y,\lambda)$ plane terminates at the upper bound.
This upper bound corresponds to the infrared `fixed point'
of Pendleton and Ross (PR)\cite{Pendleton:Ross}.

Based on these results,
we turn to the analysis of the gauged NJL model.
We make use of the compositeness condition
which relates the gauged NJL model to a gauge-Higgs-Yukawa system,
and show that
(iii)~{\it
the gauged NJL model is renormalizable
if and only if the equivalent gauge-Higgs-Yukawa system
approaches the PR fixed point in the continuum limit.
}
Furthermore,
we calculate the effective potential
of the renormalizable gauged NJL model
by using RGE technique
and explicitly perform the renormalization of it.
We find that
(iv)~{\it
the resultant expression for the effective potential
is consistent with the previous result
by KTY}\cite{KTY:PTP}, a renormalized form of
Bardeen-Love's\cite{Bardeen-Love}
potential, which was obtained in the fixed gauge coupling case
by solving SD equation in the ladder approximation.

For the nontriviality of gauge-Higgs-Yukawa systems
and hence for the renormalizability of gauged NJL model,
the presence of asymptotically free (or fixed coupling)
gauge interaction is essential.
It also turns out that
(v) {\it the asymptotic freedom of the gauge interaction
should not be too strong}.

We note that
these points (iii) and (v) were in fact suggested by
Kondo, Shuto and Yamawaki\cite{KSY}.
They calculated the decay constant $F_\pi$ using the ladder SD
equation and the Pagels-Stokar formula,
and observed that $F_\pi$ (and hence the Yukawa coupling\footnote{
Those authors and Tanabashi\cite{Yamawaki-Hungary:Tanabashi-Hiroshima}
soon recognized that
the RGE for the Yukawa coupling supplemented with compositeness
condition also leads to the same result as
$F_\pi$ obtained that way gives in the weak coupling regime.
We thank Yamawaki for informing us of this fact.
}) can be finite in the presence of
weakly-asymptotic-free gauge interaction.
Based on this observation,
they suggested that the gauge interaction might promote
the trivial NJL model to an interacting renormalizable theory.

These points (iii) and (v) were also claimed
by Krasnikov\cite{Krasnikov}
who discussed the problem in the gauged NJL model
in $4-\epsilon$ dimensions.
Although his argument based on RGE's is similar to ours and
very suggestive for the possible existence of the theory,
he discussed neither the compositeness condition
nor the relation between the gauged NJL model
and the gauge-Higgs-Yukawa system at the regularized level
with a finite cutoff $\Lambda$ (or finite $\epsilon$).
Consequently it was left unclear
how to take the continuum limit, \ie,
how to renormalize the gauged NJL model.

The paper is organized as follows.
In the next section,
we describe the nature of our approximation
and solve the RGE's for the Yukawa and quartic scalar couplings
as well as the gauge coupling
in the continuum gauge-Higgs-Yukawa system.
The solutions are characterized by some RG invariants.
In section~\ref{sec:nontrivial},
after presenting our criteria for nontriviality,
we show the above two points (i) and (ii).
The analysis of gauged NJL model is performed
in sections~\ref{sec:renorm} to \ref{sec:effpot}.
In section~\ref{sec:renorm}, we impose compositeness conditions on
the Yukawa and quartic scalar coupling constants in the \gHY\ system
to relate it to the gauged NJL model, and prove the above point (iii),
\ie, the renormalizability of the gauged NJL model.
To make the connection between the two systems complete, it is
necessary to discuss another compositeness condition on the mass
parameter.  We discuss this condition in detail in section~5
since it is important to see how the four fermion interaction $G$
determines the phase structure of the equivalent \gHY\ system.
Based on this, we calculate the effective potential of the gauged NJL
model and renormalize it explicitly in section~\ref{sec:effpot}.
The final section is devoted to conclusions.

\msection{Renormalization Group Equations
\label{sec:RGE}}

We consider the Higgs-Yukawa theory with $SU(\Nc)$ color
gauge interaction, which contains
$\Nf$ species of colored fermions $\psi_i$ ($i=1$ -- $\Nf$), each
belonging to a representation $R$ of $SU(\Nc)$, and
a color-singlet scalar field $\sigma$.
In this paper, we work in the leading order
of $1/\Nc$ expansion in the following sense.
Usually in the leading order of $1/\Nc$ expansion,
all the planar diagrams contribute
to the renormalization group equation.
However,
we are interested in the high-energy asymptotic region
where the QCD gauge coupling $g$ is small enough
to be treated perturbatively;
$\Nc g^2/4\pi \ll 1$.
We will work in the first nontrivial order
in the gauge coupling expansion.
Moreover, we consider the case where
there exist large number of fermions in the theory,
so we regards $\Nf\sim\Nc$.
However,
only $\nf$ fermions $\psi_i$ ($i=1$ -- $\nf \ll \Nf$)
among them have a degenerate large Yukawa coupling,
while the others have vanishing or negligibly
small Yukawa couplings.
As such,
the Lagrangian we consider in this paper is given by
\ba
{\cal L}_{\rm gHY} ~~ =
	 &-&\!\!\! \frac{1}{4} \left( F_{\mu\nu} \right)^2
	  +  \frac{1}{2} \left( \p_{\mu} \sigma \right)^2
	  -  \frac{1}{2} m^2 \sigma^2
	  -  \frac{\lambda}{4} \sigma^4 \nonum
	 &+&\!\!\! \sum_{i=1}^{\Nf} \wb{\psi_i}\, i\,\sD \psi_i
	  -  \sum_{i=1}^{\nf} y \sigma \wb{\psi}_i \psi_i \ ,
\label{Lag:gHY}
\ea
where
$D_\mu$ is a color-covariant derivative and
$y$ and $\lambda$ are the Yukawa and
quartic scalar coupling constants, respectively.
This choice of model is obviously inspired by
the minimal Standard Model (SM) where
$\Nc=\Nf /2=3$ and $\nf=1$.
We note, however, that
the present system crucially differs from the SM
in that the asymptotically nonfree $U(1)$ gauge interaction is
switched off.

Let $\mu_0$ denote a reference scale
at which we discuss the low-energy physics,
and $t \equiv \ln(\mu/\mu_0)$
where $\mu\,( > \Lambda_{{\rm QCD}})$ is the renormalization point.
RGE's for the gauge coupling $g(t)$, the Yukawa coupling $y(t)$
and the quartic scalar coupling $\lambda(t)$ are given
in the leading order of our approximation by
\ba
  16\pi^2 \frac{d}{dt} g(t)
&=&
  - b g^3(t) \ ,
\label{RGE:QCD}
\\
  16\pi^2 \frac{d}{dt} y(t)
&=&
  y(t) \left[ a y^2(t) - c g^2(t) \right] \ ,
\label{RGE:Yukawa}
\\
  16\pi^2 \frac{d}{dt} \lambda(t)
&=&
  u y^2(t) \left[ \lambda(t) - v y^2(t) \right] \ ,
\label{RGE:lam}
\ea
with $b$, $a$, $c$, $u$ and $v$ being positive constants:
\be
  b = \frac{11\Nc-4T(R)\Nf}{3}, \quad
  c = 6C_2(R), \quad
  a = \frac{u}{4} = 2\nf\Nc, \quad v = 1\ ,
\ee
where $C_2(R)$ is a second Casimir, and $T(R)=1/2$ and
$C_2(R)=(\Nc^2-1)/(2\Nc)$ when $R$ is the fundamental representation.
[We can also consider the fixed gauge coupling case,
by taking the limit $b \raw +0$.]
Note that in the $1/\Nc$ leading approximation,
$u=4a$ independently of the model's detail
since both $a$ and $u$ are determined solely
by the fermion one-loop contribution to the scalar self-energy.

It will be important to note that
aside from the perturbation with respect to the gauge coupling,
we are working in the leading order in $1/\Nc$ expansion
and hence neglecting the scalar loop contributions.
The latter approximation is valid as long as
the quartic scalar coupling $\lambda$ is small
compared with $\Nc y^2$:
\be
  \abs{\frac{\lambda\,}{\,y^2}}  \lsim \Nc \ .
\label{cond:Nc}
\ee
But if we follow the flow determined by
the RGE (\ref{RGE:lam}) in $1/\Nc$ leading order,
we often reach the region in which this condition breaks down.
In such a situation,
we should include the scalar loop contributions
beyond the $1/\Nc$ leading order.
{}For instance in the one-loop order,
the RGE (\ref{RGE:lam}) should be replaced by an equation of the form
\be
16\pi^2 \frac{d}{dt} \lambda(t)
  = w\lambda^2
  + u y^2(t) \left[ \lambda(t) - v y^2(t) \right]
\label{RGE:lam:1loop}
\ee
where $w>0$
is a constant of order $1$, while $u$
and $v$ are the same as before.
We shall come back to this point in the
next section.

Let us now solve the RGE's (\ref{RGE:QCD})--(\ref{RGE:lam})
and analyze the solution.
The solution to Eq.~(\ref{RGE:QCD}) for the gauge coupling
$\alpha =g^2/(4\pi)$ is well-known:
\be
\runn \equiv \frac{g^2(t)}{g^2_0}
  \equiv \frac{\alpha(t)}{\alpha_0}
  = \left( 1 + \frac{b\alpha_0}{2\pi} t \right)^{-1} \ ,
\label{sol:QCD}
\ee
where coupling constants with subscript $0$
generally denote the initial values
at the reference scale $\mu_0$;
\eg, $\alpha_0=\alpha(t\!=\!0)$.
Note that $\runs(0)=1$, and $\runn \raw +0$
in the ultraviolet (UV) limit $t \raw +\infty$.

The simplest way to solve the RGE's
(\ref{RGE:Yukawa}) and (\ref{RGE:lam})
is to observe that the quantities
\ba
  h(t)
&\equiv&
  - \run{-1+\cob}
  \left[ 1 - \frac{c-b}{a}\frac{g^2(t)}{y^2(t)} \right] \ ,
\label{def:h}
\\
  k(t)
&\equiv&
  - \run{-1+2\cob}
  \left[1 - \frac{2c-b}{2av}
  \frac{\lambda(t)}{y^2(t)}\frac{g^2(t)}{y^2(t)} \right]
\label{def:k}
\ea
are RG invariant:
$0 = (d/dt) h(t) = (d/dt) k(t)$.
Then the solutions of the RGE's are specified
once the values of the RG invariants $h$ and $k$ are given.
Let $h_0$ and $k_0$ denote the values of the invariants.
Then we can write the solutions as
\ba
y^2(t)  &=& \frac{c-b}{a} g^2(t)
  \left[1 + \hz \run{1-\cob} \right]^{-1} \ ,
\label{sol:Yukawa}
\\
\lambda(t)
  &=& \frac{2av}{2c-b} \frac{y^4(t)}{g^2(t)}
  \left[1 + \kz \run{1-2\cob} \right] \ .
\label{sol:lam}
\ea

Before entering the detailed analysis in the next section,
let us briefly see how the solutions look like in some limiting cases.
First,
the fixed gauge coupling limit $b \raw +0$
can be taken by using
\be
 \run{-\cob}  \mathop{\lra}_{b \raw 0}
\exp{\left(\frac{\alpha}{\alc} t\right)}
    = \left(\frac{\mu\,}{\,\mu_0}\right)^{\aoac} ,
\label{fixlimit}
\ee
where $\alpha\,(=\alz)$ is the fixed gauge coupling constant and
\be
  \frac{1}{\alc} \equiv \frac{c}{2\pi} = \frac{3C_2(R)}{\pi} \ .
\label{def:alc}
\ee
Then we obtain from Eqs.~(\ref{sol:Yukawa}) and
(\ref{sol:lam})
\ba
y^2(t)
  &=&	\frac{(4\pi)^2}{2a} \frac{\alpha}{\alc}
	\left[ 1 + \hz \exp\left(\frac{\alpha}{\alc} t \right)
\right]^{-1} \ ,
\label{sol:Yukawa:fix}
\\
\lambda(t)
  &=&	\frac{2av}{(4\pi)^2} \frac{\alc}{\alpha} y^4(t)
	\left[ 1 + \kz \exp\left(\frac{2\alpha}{\alc} t\right) \right]
\label{sol:lam:fix} \ .
\ea
Next we consider the limiting case $b=c$.
Although the Yukawa coupling (\ref{sol:Yukawa})
appears to vanish in this case,
a careful consideration of RGE
leads to a non-vanishing result.
The correct way is first to write the RG invariant as
\be
\hz = -1 + \frac{c-b}{b} h_1
\label{def:h1}
\ee
and then to take the limit $c \raw b$ in Eq.~(\ref{sol:Yukawa}).
Then we find that
\be
y^2(t)	 = \frac{b}{a} g^2(t)
	 \left[ h_1 + \ln \runn \right]^{-1} \ .
\label{sol2:Yukawa}
\ee
Note that $h_1=(b/a)(g^2_0/y^2_0)$ is RG invariant
in this limiting case.
The case $b=2c$ can be treated in a similar manner.

\msection{Nontriviality of Gauge-Higgs-Yukawa System
\label{sec:nontrivial}}

We now use the solutions of the RGE's
found in the preceding section
to investigate under what circumstances
the gauge-Higgs-Yukawa system is nontrivial.
We adopt the following criteria
for the nontriviality:
all the running coupling constants
should not diverge at any finite $t$($>0$),
should not vanish identically
and should not violate the consistency of the theory,
such as the unitarity and vacuum stability.
Note that we are demanding the nontriviality
as a gauge-Higgs-Yukawa system.
Namely, for instance, we do not call nontrivial
the system with $y(t)\equiv\lambda(t)\equiv0$,
which is in fact a nontrivial QCD-like theory
with the scalar field decoupled completely.
Note also that the mass parameter poses no problem,
since it is multiplicatively renormalized in the $\MS$ scheme.
Even if the multiplicative factor diverges at a finite $t$,
we can avoid the problem by setting $m^2\equiv0$ and it does not imply
the triviality of the theory.

We should remark that we are restricting ourselves only to the cases
where the gauge coupling remains small in the UV region.
If the Yukawa or quartic scalar couplings become so large
that the $1/\Nc$ expansion breaks down,
they might affect the behavior of $g^2(t)$ substantially such that
our perturbation assumption with respect to $g^2(t)$ is violated.
So, logically, there remain possibilities that
all the couplings become large but have a nontrivial UV fixed point
so that the system gives a nontrivial theory.
This is outside the scope of this paper.

\subsection{Fixed Point Solution}

It is instructive to analyze first the case
in which the RG invariants are $\hz=\kz=0$.
In this case,
the solutions (\ref{sol:Yukawa}) and (\ref{sol:lam})
reduce to
\ba
\frac{y^2(t)}{g^2(t)}
	&=& \frac{c-b}{a} \ ,
\label{sol:PR:Yukawa}
\\
\frac{\lambda(t)}{y^2(t)}
	&=& \frac{2av}{2c-b} \frac{y^2(t)}{g^2(t)}
	 =  \frac{2v\,(c-b)}{2c-b} \ .
\label{sol:PR:lam}
\ea
We see that
the behavior of the Yukawa and quartic scalar couplings,
$y(t)$ and $\lambda(t)$, are completely determined
by that of the gauge coupling $g(t)$.
This corresponds to the ``coupling constant reduction''
in the sense of Kubo, Sibold and Zimmermann\cite{KSZ:reduction}.
In the context of RGE, it corresponds to
the Pendleton-Ross (PR) fixed point\cite{Pendleton:Ross}.\footnote{
For earlier works on RG fixed points, see Refs.~\cite{Chang:Ma,GW:CEL}.}
This can be seen by rewriting the RGE's
(\ref{RGE:QCD})--(\ref{RGE:lam}) into
\ba
8\pi^2 \frac{d}{dt} \left[ \frac{y^2(t)}{g^2(t)} \right]
	&=& ag^2(t) \left[ \frac{y^2(t)}{g^2(t)} \right]
	\left[ \frac{y^2(t)}{g^2(t)} - \frac{c-b}{a} \right] \ ,
\label{RGE:PR:Yukawa}
\\
8\pi^2 \frac{d}{dt}
	\left[ \frac{g^2(t)}{y^2(t)}\frac{\lambda(t)}{y^2(t)}
	\right]
	&=& (2c-b)g^2(t)
	\left[ \frac{g^2(t)}{y^2(t)}\frac{\lambda(t)}{y^2(t)}
	- \frac{2av}{2c-b} \right] \ .
\label{RGE:PR:lam}
\ea
Clearly,
the solution $\hz=0$ and $\kz=0$
is the fixed point of the RGE's
(\ref{RGE:PR:Yukawa}) and (\ref{RGE:PR:lam}).
[Note that the former RGE has two fixed points,
\ie, $y^2/g^2=0$ and the PR fixed point $y^2/g^2=(c-b)/a$.]

Observe that
the expressions (\ref{sol:PR:Yukawa}) and (\ref{sol:PR:lam})
make sense only when $c \geq b$.
[Otherwise,
the Yukawa coupling becomes complex
implying the violation of unitarity,
or the quartic scalar coupling becomes negative
implying the instability of the vacuum.]
When $c > b$,
the Yukawa and quartic scalar couplings
as well as the gauge coupling are asymptotically free.
This is a sufficient condition
for the nontriviality of this system.
[The case $c=b$ requires some caution
and will be discussed later.]

We thus see that if $c>b$,
the gauge-Higgs-Yukawa system with $\hz=\kz=0$
corresponds to the PR fixed point solution to the RGE's
and hence a nontrivial theory.
This consideration also clarifies
the meaning of the RG invariants $\hz$ and $\kz$;
they characterize how much the RG flow deviates from the PR one.
Then the next question will be
how about the case $\hz \neq 0$ or $\kz \neq 0$.
We will systematically explore this in the following.

\subsection{Yukawa Coupling}

Let us first discuss the Yukawa coupling $y(t)$.
As we have seen in the preceding subsection,
the behavior of the Yukawa coupling will be quite different
depending on whether $c>b$ or $c<b$.
Also the case $c=b$ requires a special care.
Moreover, the sign of the RG invariant $\hz$
will be important.\footnote{
The RG invariant $\hz$ should be a finite constant
since otherwise,
the Yukawa coupling vanishes identically
and the theory becomes trivial as a gauge-Higgs-Yukawa system.
}
We proceed in the order i) $c>b$, ii) $c<b$, iii) $c=b$.
It will turns out that
the Yukawa interaction can be nontrivial
only for the first case $c>b$.

\subsubsection{$c>b$ case}

Let us start with the case $c>b$.
Since $\run{1-c/b}$ becomes large
as $t \raw +\infty$,
the UV asymptotic behavior of $y^2(t)$ in Eq.~(\ref{sol:Yukawa})
is given by
\be
y^2(t)	\sim \frac{c-b}{a} \frac{g^2_0}{\hz} \run{\cob}
	\lra \pm 0 \ .
\label{UV1:Yukawa}
\ee
Then it appears that
the Yukawa coupling is asymptotically free
and the theory is nontrivial in this case.
However,
the sign of the RG invariant $\hz$ is crucial
in order that this is really true.

{}For $0 < \hz < +\infty$,
we see from the solution (\ref{sol:Yukawa}) that
$y^2(t)$ remains positive and finite.
Then the UV asymptotic behavior is really given
by Eq.~(\ref{UV1:Yukawa}), $y^2(t) \raw +0$,
and the theory is nontrivial.

{}For $-\infty < \hz < 0$, however,
this is not the case.
First of all, notice that
we should exclude $\hz \leq -1$ from the beginning
in order that $y^2_0 \equiv y^2(t\!=\!0)$ is finite and positive.
Even for $-1 < \hz < 0$,
there exists a Landau pole at a finite value of $t>0$
satisfying
\be
1 + \frac{b\alz}{2\pi} t
	= \left( \frac{-1}{\hz} \right)^{b/(c-b)}
\label{poleposition}
\ee
at which $y^2(t)$ diverges and changes its sign.
So even if $y^2_0$ is arranged to be positive and finite
at an infrared (IR) scale $\mu_0$,
such a property is not preserved in the UV region.
The asymptotic behavior (\ref{UV1:Yukawa}) holds
in the UV region beyond the Landau pole,
but the unitarity is violated there.
Thus the theory is trivial according to our criteria.

The RG flows in the $(\alpha,y^2)$ plane are shown in
Fig.~\ref{fig:alpha:Yukawa}.

\figinsert{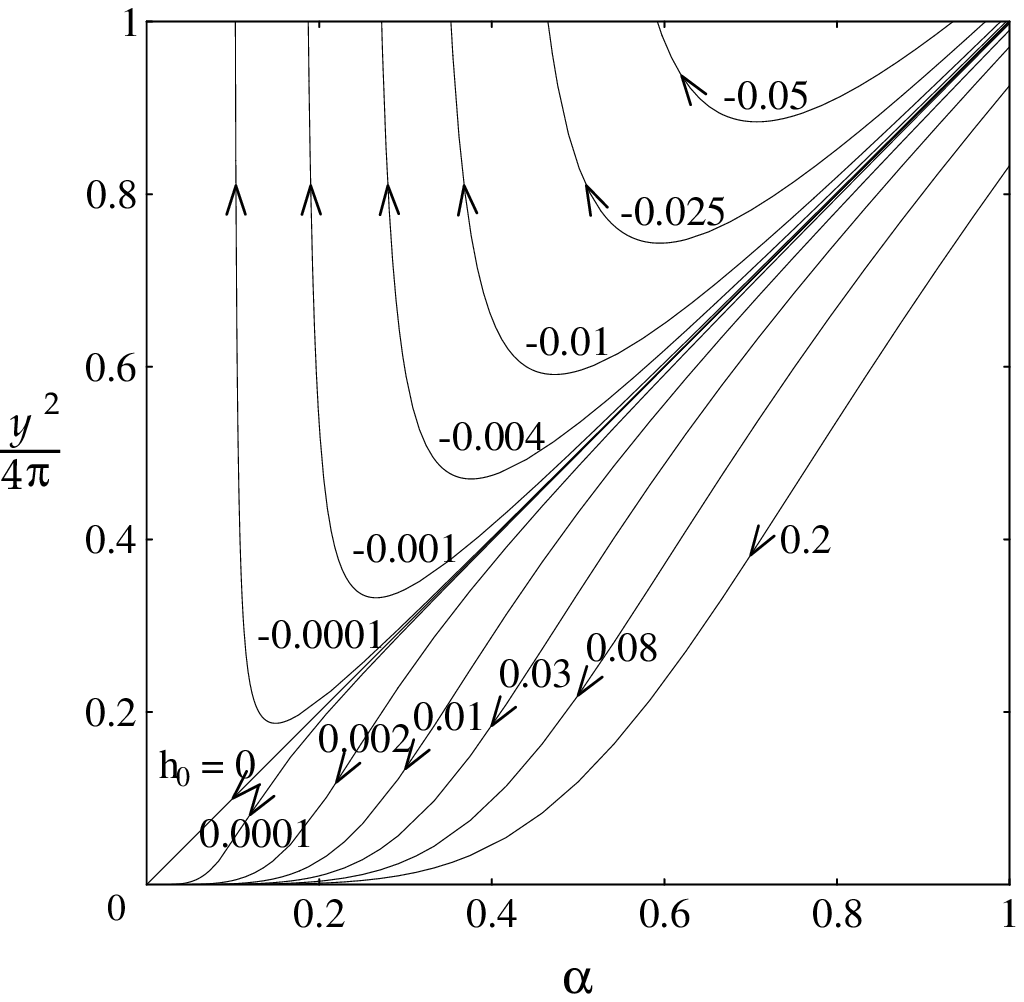}{
RG flows in the $(\alpha, y^2)$ plane. The arrows denote the flow
directions {\it toward} the UV region. The flows for $h_0 \geq 0$ below the
$h_0 = 0$ line all converge to the origin (\ie, asymptotically free)
while those for $h_0 <0$ diverge at finite $t$ (or $\alpha$).
}{fig:alpha:Yukawa}

In this way we conclude that
the Yukawa interaction with $c>b$ is nontrivial
if and only if $\hz \geq 0$.
Note that
the nontriviality
\footnote{
The IR limit of $y(t)$ is given by the fixed point value
(\ref{sol:PR:Yukawa}),
but such IR limit can not be treated
by the present perturbative calculation
with respect to the gauge coupling.
}
does not necessarily imply the coupling reduction:
there exist a one parameter family of theories ($0 < \hz < +\infty$)
which do {\it not} correspond to the PR fixed point ($\hz=0$).
Indeed,
the value of the Yukawa coupling $y^2_0$
at the IR scale $\mu_0$ is given by
\be
0 \ < \ \  y^2_0 = \frac{1}{1+\hz} \frac{c-b}{a} g^2_0
	  \ \  \leq \ \  \frac{c-b}{a}g^2_0
	  \equiv y^2_{\star} \ .
\label{suffcond}
\ee
We see that
the low-energy Yukawa coupling $y^2_0$
can take any value in between $0$ and $y^2_{\star}$,
and that
the upper bound $y^2_{\star}$ is given
by the PR fixed point (\ref{sol:PR:Yukawa}).

\subsubsection{$c<b$ case}

In this case of $c<b$,
$\run{1-\cob} \raw +0$, and
the UV asymptotic behavior of the solution (\ref{sol:Yukawa})
is given, instead of (\ref{UV1:Yukawa}), by
\be
y^2(t)	\ \sim \ - \frac{b-c}{a} g^2(t)
	\ \lra \ - 0 \ .
\label{UV2:Yukawa}
\ee
Since the RHS of this equation is negative,
there is no chance for us to have a nontrivial theory in this case.

Indeed,
for $0 \leq \hz < +\infty$,
$y^2(t)$ is always negative,
and the unitarity is violated
although there is no Landau pole.
{}For $-\infty < \hz <0$,
we meet a Landau pole;
even if $y^2_0$ is arranged to be finite and positive,
$y^2(t)$ diverges at finite $t$.

It would be amusing, however, to note that
this case $c<b$ can be related to the above case of $c>b$
by the following `symmetry' of the RGE's under the transformation
$(t,y,b,c) \raw (-t,iy,-b,-c)$
with others inert.
This exchanges the UV with the IR region,
and $c>b$ with $c<b$ case.
We then find that
the nontrivial theory with $c>b$ and $\hz \geq 0$
in the physical region $y^2>0$
is mapped to the theory with $c<b$ and $\hz \geq 0$
in the unphysical region $y^2<0$.

\subsubsection{$c=b$ case}

In this case,
we should use the solution (\ref{sol2:Yukawa})
with $h_1$ given in Eq.~(\ref{def:h1}).
Since $\ln \runn$
behaves like $-\ln t$ as $t \raw + \infty$,
the UV asymptotic behavior is
\be
y^2(t)	= \frac{b}{a} g^2(t)
	  \left[ h_1 - \ln \left( 1 + \frac{b\alz}{2\pi}t \right)
	  \right]^{-1}
	\ \longrightarrow \ - \frac{b}{a} \frac{g^2(t)}{\ln t} \ .
\ee
Then as in the case $c<b$, we find that
the Yukawa coupling $y^2(t)$ is always negative for $h_1 \leq 0$,
and that for $h_1>0$
it always has a Landau pole at $\ln(1+b\alz t/2\pi)=h_1$.

Combined with the result for the case $c<b$,
we conclude that
the Yukawa interaction with $c \leq b$ is necessarily trivial.

\subsubsection{Further Comments}

It is well known that
the Yukawa interaction is trivial by itself
since the RGE (\ref{RGE:Yukawa}) without $g^2$
is easily solved to produce a Landau pole.
{}For the existence of the nontrivial Yukawa interaction,
the presence of asymptotically free
(or fixed coupling) gauge interaction,
$b\geq0$, is crucial.
Our result, however, shows that $b$ should be smaller than $c$;
the asymptotic freedom of gauge interaction
should not be too strong.
Let us now think about the reason.

{}For a moment,
let us go back to the original RGE (\ref{RGE:Yukawa}).
Based on this equation,
one might expect that
the Yukawa interaction would be nontrivial if
\be
y^2_0 \leq \frac{c}{a} g^2_0 \ ,
\label{nececond}
\ee
since the RGE (\ref{RGE:Yukawa}) has
an IR-stable `fixed point' $y^2=(c/a)g^2$ of order $O(g^2)$;
If this condition is satisfied at some scale,
the Yukawa coupling has negative beta function
and seems asymptotically free.
The point is, however, that
the `fixed point' itself moves to zero as $t$ becomes large,
and it will `overtake' the Yukawa coupling $y^2(t)$
when the asymptotic freedom of gauge interaction is too strong.
The condition (\ref{nececond}) is necessary,
but not sufficient
in order for the theory to give really an asymptotically free theory.
The argument based on the exact solution
(\ref{sol:Yukawa}) shows that
the sufficient condition is given by Eq.~(\ref{suffcond}).
This in turn implies that
$b$ should be positive, but not too much.

This fact was originally pointed out by Krasnikov\cite{Krasnikov}
from the analysis of the RGE's in the one-loop order.
His conclusion was that
there exists a minimal number for the colored fermions
which weaken the asymptotic freedom of QCD gauge interaction.
In our case with RGE's in the $1/\Nc$-leading order,
we have modified the $1/\Nc$ expansion
by including the $\Nf \sim \Nc$ species of fermions.
For instance, if $\Nf \simeq 2\Nc$ and
$R$ is the fundamental representation, then $c\simeq3\Nc$ and
$b\simeq(7/3)\Nc$, and hence the condition $c>b>0$ is
satisfied.

Finally it is interesting to note that
the Standard Model
with three generations of fermions
(with $SU(2)_{\rm L}\times U(1)_{\rm Y}$ gauge interactions
switched off) just gives an example of this category of theories,
so that it would nontrivially exist
even when the cutoff goes to infinity.

\subsection{Quartic Scalar Coupling}

We now proceed to analyze the quartic scalar coupling $\lambda(t)$.
We restrict ourselves to the case of $c>b$ and $\hz\geq0$
in which a nontrivial limit of Yukawa coupling can be taken.

The solution (\ref{sol:lam})  explicitly reads
\be
\lambda(t)
	= \frac{2v\,(c-b)^2}{a\,(2c-b)} g^2_0 \runn
	\frac{ 1 + \kz \run{1-2\cob} }
	     {\left[ 1 + \hz \run{1-\cob} \right]^2} \ .
\label{sol:lam:2}
\ee
When $\kz=0$, the asymptotic behavior in the UV limit
$\run{1-2c/b} \raw +\infty$ is given by
\be
\renewcommand{\arraystretch}{1.5}
\lambda(t) \ \mathop{\sim}_{\kz=0}\
	\frac{2v\,(c-b)^2}{a\,(2c-b)} g^2_0 \times
\left\{
  \begin{array}{ll}
    \displaystyle
	\runn
      & {\rm for } \quad \hz=0
  \\
    \displaystyle
      \frac{1}{(\hz)^2\run{1-2\cob}}
      & {\rm for } \quad \hz>0
  \end{array}
\right. \ ,
\label{UV2:lam:k=0}
\ee
which shows that $\lambda(t)$ is asymptotically free in this case.
On the other hand for $\kz\neq0$ case,
the UV asymptotic behavior of
Eq.~(\ref{sol:lam:2}) becomes
\be
\renewcommand{\arraystretch}{1.5}
\lambda(t) \sim
	\kz \frac{2v\,(c-b)^2}{a\,(2c-b)} g^2_0 \times
\left\{
  \begin{array}{ll}
    \displaystyle
	\run{2\left(1-\cob\right)}
      & {\rm for } \quad \hz=0
  \\
    \displaystyle
      \frac{1}{(\hz)^2}
      & {\rm for } \quad \hz>0
  \end{array}
\right. \ ,
\label{UV2:lam}
\ee
which implies that $\kz<0$ case is excluded since $\lambda(t)$ becomes
negative while $\lambda(t)$ remains finite and positive at any finite
$t$ when $\kz>0$.
{}From these we are tempted to conclude that
the system gives a nontrivial theory when $\kz \geq 0$.
However, this conclusion is premature.
As we see shortly,
the conclusion for the case $\kz = 0$ remains true,
but the cases $\kz > 0$ and $\kz < 0$ should be reconsidered more
carefully.

Let us recall that
we are working in the leading order in the $1/\Nc$ expansion
and neglecting the scalar loop contributions.
As mentioned before,
this approximation is valid only when
the condition (\ref{cond:Nc}) is satisfied.
Actually, from Eqs.~(\ref{sol:Yukawa}) and (\ref{sol:lam}),
our solution for $\kz\neq0$ is such that
\ba
\renewcommand{\arraystretch}{1.5}
\frac{\lambda(t)}{y^2(t)}
	& = & \frac{2v (c-b)}{2c-b}
	\frac{ 1 + \kz \run{1-2\cob} }
	     { 1 + \hz \run{1- \cob} }
\label{RYLam:lam/yy} \\
     & \displaystyle \mathop{\longrightarrow}_{t\gg 1} &
\kz \frac{2v\,(c-b)}{2c-b} \times
\left\{
  \begin{array}{ll}
   \displaystyle
	\run{1-2\cob}
       & {\rm for} \quad \hz=0
  \\
   \displaystyle
        \frac{1}{\hz}
        \run{-\cob}
       &{\rm for} \quad \hz > 0
  \end{array}
\right. \ .
\label{UV2:lam/yy}
\ea
We see that
independently of whether $\hz=0$ or $\hz>0$,
the validity condition (\ref{cond:Nc})
for the $1/\Nc$ expansion breaks down unless $\kz=0$.
The conclusion for $\kz=0$ is not altered since
\be
0 \  <  \ \ \frac{\lambda(t)}{y^2(t)} \  \
	\leq \ \ 2v \frac{c-b}{2c-b} \
	\simeq  O\left( 1 \right)
	\qquad 	(\,\hz \geq 0 \,) \ .
\ee

In the case of $\kz \neq 0$, therefore,
we should include the scalar loop contributions
into the RGE (\ref{RGE:lam}), as mentioned before.
When $\kz > 0$,
Eq.~(\ref{UV2:lam/yy}) shows that we enters the region,
\be
  \lambda(t)
	\quad  \gsim \quad   \Nc \, y^2(t) \,
	\quad \big( \, \sim   \Nc \, g^2(t) \, \big) \ ,
\ee
at a certain UV scale $t$.
Since the gauge coupling $g^2(t)$ and
Yukawa coupling $y^2(t)$ are both small there,
the RGE to be used will take the form
\be
\frac{d}{dt} \lambda(t)
	= \beta_{\lambda}\left(\lambda(t)\right)
 	+ O\left( y^2(t), \, g^2(t) y^2(t) \right) \ ,
\label{RGE:lam:pure}
\ee
where $\beta_{\lambda}(\lambda)$ is the beta function
of the pure $\lambda\phi^4$ theory.
Then we can apply ordinary discussion for
triviality of pure $\lambda\phi^4$ theory:
$\lambda(t)$ will hit a Landau pole,
instead of having a finite limit (\ref{UV2:lam}).
We thus conclude that
the theory is trivial in this case of $\kz > 0$.

When $\kz < 0$, on the other hand,
$\lambda(t)$ becomes negative already in the region
in which the $1/\Nc$ expansion is valid.
In order for the system to give a nontrivial theory,
the only possibility is that
$\lambda(t)$ turns positive at a certain finite $t$
owing to the scalar loop contributions.
But this possibility is actually excluded as follows:
if this happens,
$\vert \lambda(t) \vert$ becomes small around the zero
and the behavior is well described there
by the one-loop RGE (\ref{RGE:lam:1loop}),
\be
16\pi^2 \frac{d}{dt} \lambda(t)
  = w\lambda^2
  + u y^2(t) \left[ \lambda(t) - v y^2(t) \right] \ .
\end{equation}
This equation tells us that $\lambda(t) \raw -0$ as $t \raw \infty$
and hence that $\lambda(t)$ remains negative at any finite $t$.
The stability of the system is not restored
in this case of $\kz < 0$.

The RG flows in the $(\lambda, y^2)$ plane are exemplified for the
case $h_0=1, c\!=\!5 > b\!=\!1$ $(a=4)$ in
Fig.~\ref{fig:lambda:Yukawa}.
\figinsert{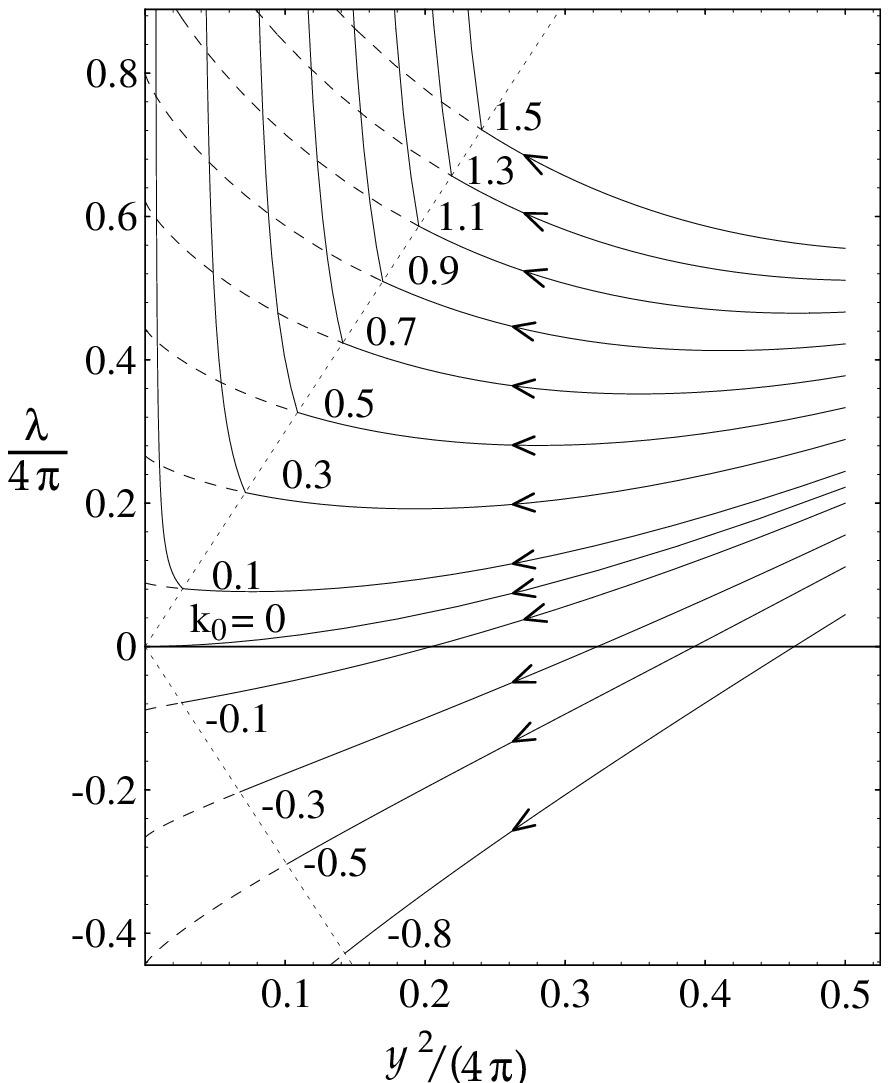}{
RG flows in the $(\lambda, y^2)$ plane with gauge-coupling
$\alpha(t\!=\!0)$ fixed to be 1.  The dotted line extending from the
origin denotes the validity bound ({\ref{cond:Nc}}) for the $1/\Nc$
leading order approximation, beyond which we have drawn the flows for
$k_0>0$ by using the one-loop $\beta_\lambda(\lambda)$ of the pure
$\lambda\phi^4$ theory.}{fig:lambda:Yukawa}

Our conclusion here is that
when $c>b$ and $\hz \geq 0$ for which
the Yukawa interaction is nontrivial,
{\it the quartic scalar interaction is also nontrivial
if and only if $\kz=0$}.
This means that for the nontrivial gauge-Higgs-Yukawa system,
the quartic scalar coupling $\lambda(t)$ is always reduced to
\be
\lambda(t)
 = \frac{2av}{2c-b} \frac{y^4(t)}{g^2(t)} \ .
\ee
Since this is $t$-independent, it also holds
for the low-energy couplings at the scale $\mu_0$:
\be
\lambda_0
	\equiv  \lambda(t\!=\!0)
	= 	\frac{2av}{2c-b} \frac{y^4_0}{g^2_0} \ .
\label{triviality}
\ee
{\it One combination of the Yukawa and the quartic scalar
interaction operators is irrelevant in the nontrivial
gauge-Higgs-Yukawa system. }
The relation (\ref{triviality})
between $\lambda_0$ and $y_0^2$ is depicted
by the bold line in Fig.~3.
\figinsert{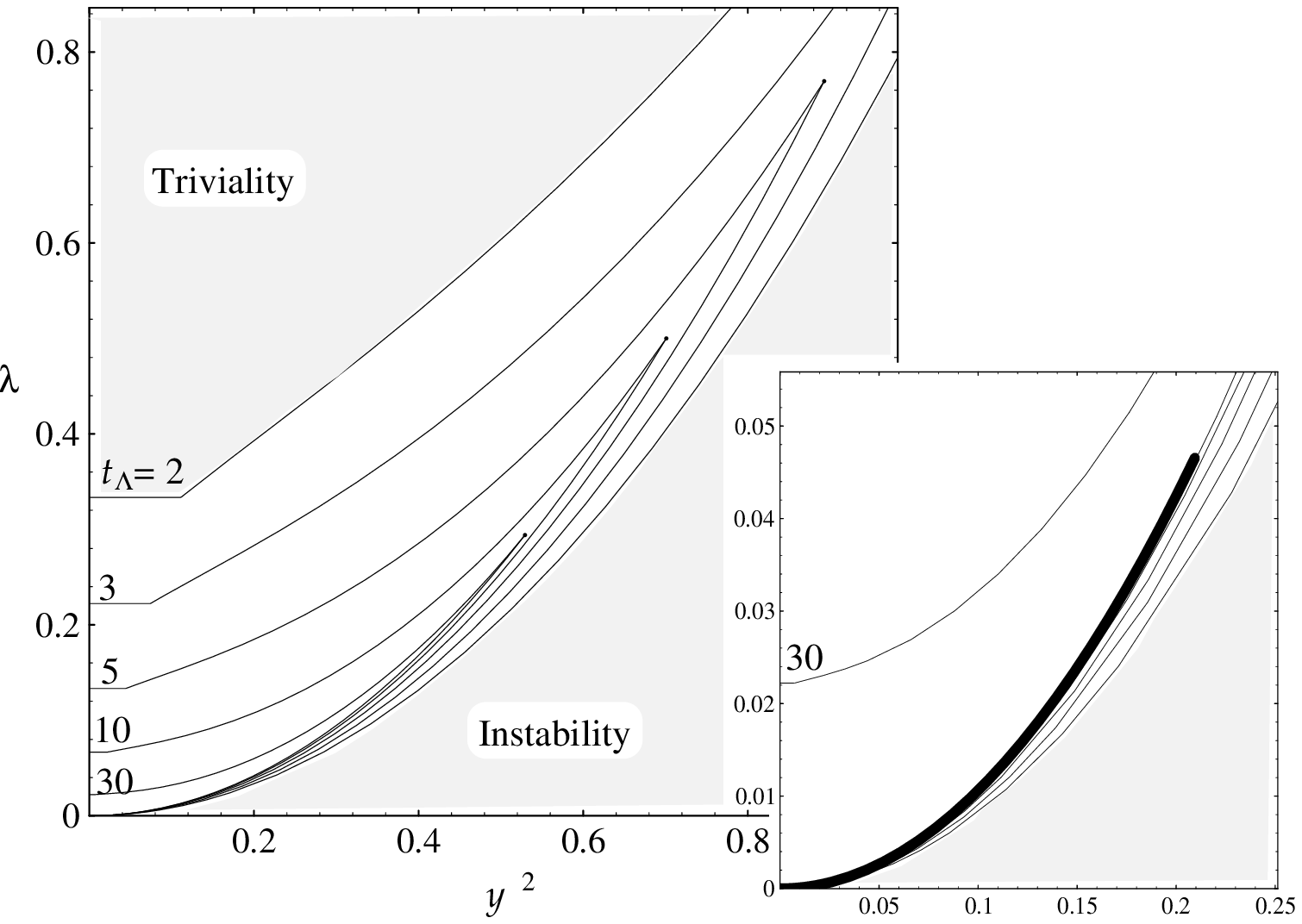}{
Regions in the $(\lambda_0, y_0^2)$ plane restricted by triviality and
instability bounds. The reference scale $\mu_0$ of $t=\ln(\mu/\mu_0)$
is fixed by $\alpha(t\!=\!0)=0.1$, and the parameters $\Nc=3, \Nf=6,
n_{\rm f}=1$ and $R=\;$fundamental repr.
are taken. The triviality bounds (upper lines) determined
by $\lambda(t_\Lambda)=\infty$ and the instability bounds by
$\lambda(t_\Lambda)=0$ are shown for $t_\Lambda=2, 3, 5, 10, 30$
and $\infty$. The allowed regions shrink to the line
(\ref{triviality}) as $t_\Lambda\rightarrow \infty$, which is drawn by
the bold line in the enlarged figure in the bottom right.
Again the running of $\lambda(t)$ is determined by using one-loop
$\beta_\lambda(\lambda)$ of pure $\lambda\phi^4$ theory
whenever the inequality $\lambda(t)\ge\Nc y^2(t)$ holds.
}{fig:triviality}
In the SM,
this is related to the so-called ``triviality bound''
for the masses of Higgs boson and top quark\cite{CMPP:pert.unif.}.
What is special to the present case is that
the allowed parameter region for $y^2_0$ and $\lambda_0$
shrinks to a line
connecting the ``Gaussian" and PR fixed points.
Actually, in Fig.~\ref{fig:triviality}, we have also drawn
``triviality bound'' and ``instability bound'' on the
$(y_0^2,\lambda_0)$ plane, which are determined by requiring
$\lambda(\tcut)=\infty$ and $\lambda(\tcut)=0$, respectively, at a
certain UV scale $\tcut$.
The theories with the parameters $(y_0^2,\lambda_0)$ inside the region
enclosed by those two bounds give allowed cutoff theories possessing
finite and positive couplings $y^2(t)$ and $\lambda(t)$ over the
energy region $0\leq t<\tcut$.
As $\tcut$ goes to infinity, this allowed parameter region is indeed
seen to shrink to the line given by Eq.~(\ref{triviality}).

\msection{Renormalizability of Gauged NJL Model
\label{sec:renorm}}

We now turn to the discussion on the renormalizability of
the gauged NJL model.
The Lagrangian of the model is given by
\be
{\cal L}_{\rm GNJL}
	= - \frac{1}{4} \left( F_{\mu\nu} \right)^2
	  + \sum_{i=1}^{\Nf} \wb{\psi_i}\, i\,\sD \psi_i
	  + G \sum_{i=1}^{\nf} \left( \wb{\psi_i} \psi_i \right)^2 \ ,
\label{GNJL}
\ee
where $G$ is a four-fermion coupling constant\rlap.\footnote
{In this paper,
we consider only the attractive four-fermion interaction, $G>0$.
The renormalizability of the gauged NJL models
with the repulsive four-fermion interaction
is beyond the scope of this paper.
}
For later convenience,
we introduce a dimensionless coupling constant $g_4$ by
\be
  G \equiv \frac{(4\pi)^2}{a} \frac{g_4(\Lambda)}{\Lambda^2} \ ,
\ee
where $\Lambda$ is the UV cutoff.

As usual,
we rewrite the four-fermion interaction by the standard method
of introducing an auxiliary field.
The resultant Lagrangian can be identified with
that of a gauge-Higgs-Yukawa system
under a suitable compositeness condition,
which in particular requires\cite{CompCond} the vanishingness of the
wave function renormalization factor,
$Z_{\sigma}=0$,
of the scalar field $\sigma$.
According to Bardeen, Hill and Lindner\cite{BHL},
such compositeness condition can be stated
most conveniently as a set of boundary conditions
for the running parameters of the gauge-Higgs-Yukawa system
at the compositeness scale $\tcut \equiv \ln(\Lambda/\mu_0)$,
\ie, at the cutoff of the gauged NJL model.
As for the dimensionless coupling constants\cite{BHL}, it reads
\ba
  \frac{1}{y^2(\tcut)} &=& 0 \ ,
\label{compcond:Yukawa}
\\
  \frac{\lambda(\tcut)}{y^4(\tcut)} &=& 0 \ .
\label{compcond:lam}
\ea

In addition to these, the compositeness condition
gives a boundary condition for the mass
parameter $m^2$ of $\sigma$, which relate $m^2(t_\Lambda )$
to the four-fermion coupling constant $G$.
There are, however,
some subtleties in this condition for the mass parameter.
So we
defer the discussion to the next section.

Now, with the compositeness condition
(\ref{compcond:Yukawa}) and (\ref{compcond:lam}),
we discuss
the renormalizability of the gauged NJL model
by utilizing the result
in section~\ref{sec:nontrivial} for
the nontriviality
of the gauge-Higgs-Yukawa system.
We shall show in the next section that the mass parameter
$m^2(t)$ can be made finite by
the renormalization of the four-fermion coupling $G$.
So we concentrate on the renormalization problem of the coupling
constants $y^2$ and $\lambda$ here.
The conditions (\ref{compcond:Yukawa}) and (\ref{compcond:lam})
uniquely specify the RG flow
corresponding to the gauged NJL model;
by substituting them into Eqs.
(\ref{def:h}) and (\ref{def:k}),
the RG invariant constants $\hz$ and $\kz$ are determined
in terms of gauge coupling constant at the cutoff scale as
\ba
\hz 	&=& - \eta^{\cob-1}(\tcut)
	 =  - \left(\frac{\alpha(\tcut)}{\alpha_0}
	      \right)^{\cob-1} \ ,
\label{h0:compcond}
\\
\kz	&=& - \eta^{2\cob-1}(\tcut)
	 =  - \left(\frac{\alpha(\tcut)}{\alpha_0}
	      \right)^{2\cob-1} \ .
\label{k0:compcond}
\ea
Then the Yukawa and quartic scalar couplings
are given by\footnote{
It is easy to check that the validity condition (\ref{cond:Nc}) is
satisfied for $t<\tcut$.}
\ba
y^2(t) &=&
	\frac{c-b}{a} g^2(t)
	\left[ 1 - \left( \frac{\alpha(t)}{\alpha(\tcut)}
		   \right)^{1-\cob}
	\right]^{-1}
    \ \equiv \
    y_\Lambda^2(t) \ ,
    \qquad
    \raisebox{-20pt}[0pt][0pt]{$\displaystyle (t<\tcut)$}
\label{sol:comp:Yukawa}
\\
  \frac{\lambda(t)}{y^4(t)}
&=&
  \frac{2av}{2c-b} \frac{1}{g^2(t)}
  \left[ 1 - \left( \frac{\alpha(t)}{\alpha(\tcut)}
	   \right)^{1-2\cob}
  \right]
  \ \equiv \  \frac{\lambda_\Lambda(t)}{y_\Lambda^4(t)}
\ .
\label{sol:comp:lam}
\ea
The gauged NJL model with a cutoff $\Lambda$
is equivalent to the gauge-Higgs-Yukawa system
with the same cutoff $\Lambda$
with the Yukawa and quartic scalar couplings given by
Eqs.~(\ref{sol:comp:Yukawa}) and (\ref{sol:comp:lam}).

When we vary the cutoff $\Lambda$,
the equivalent gauge-Higgs-Yukawa systems specified by
Eqs.~(\ref{h0:compcond}) and (\ref{k0:compcond}) define
{\it a sequence of theories parameterized by $\Lambda$}.
Now the question of the renormalizability
of the gauged NJL model is whether the sequence converges
in the $\Lambda\raw\infty$ limit to a nontrivial theory.
On the other hand, we know already that
the gauge-Higgs-Yukawa system is nontrivial
only when $\hz \geq 0$ and $\kz=0$.
In view of the expressions
(\ref{h0:compcond}) and (\ref{k0:compcond}),
the only possibility is that
\be
\hz \ , \ \kz \lra - 0 \ .
\label{limit:h0k0}
\ee
This is realized only when $c>b$,
which is just the condition for the nontriviality
of gauge-Higgs-Yukawa system.

We therefore conclude that
the gauged NJL model becomes renormalizable
(without becoming trivial theory)
if and only if its matter content satisfies $c>b$\footnote{
This condition corresponds to $A>1$ of Kondo, Shuto and
Yamawaki's\cite{KSY},
which is required for the finiteness of the decay constant $F_\pi$.
},
and that
the resultant continuum theory becomes
identical with the gauge-Higgs-Yukawa system
just on the PR fixed point, $\hz=\kz=0$.
The resultant Yukawa and quartic scalar couplings
are given by their fixed point values
(\ref{sol:PR:Yukawa}) and (\ref{sol:PR:lam}).

Note that the condition $c>b$ is automatically satisfied
when we consider the fixed coupling limit $b \rightarrow +0$
since $c$ is a positive constant.
It then follows that
the gauged NJL model in this limit is always renormalizable.
This agrees with KTY's result for the fixed gauge coupling
case.\cite{KTY:PTP}
The renormalized Yukawa and quartic scalar couplings
(\ref{sol:comp:Yukawa}) and (\ref{sol:comp:lam})
under the compositeness condition are found
in the limit $b\rightarrow+0$
by using the formula (\ref{fixlimit}) to reduce
to
\ba
y_\Lambda^2(t)
	&=& \frac{(4\pi)^2}{2a} \frac{\alpha}{\alc}
	\left[ 1 - \left(\frac{\mu}{\Lambda}\right)^{\aoac}
	\right]^{-1}
        \
        \mathop{-\!\!\!-\!\!\!-\!\!\!-%
        \!\!\!\!\!\longrightarrow}%
        _{\Lambda \raw \infty}
        \
        \frac{(4\pi)^2}{2a} \frac{\alpha}{\alc}
	\equiv
        y^2_{\star}
        \ ,
\label{sol:comp:Yukawa:fix}
\\
\frac{\lambda_\Lambda(t)}{y_\Lambda^4(t)}
	&=& \frac{2av}{(4\pi)^2} \frac{\alc}{\alpha}
	\left[ 1 - \left(\frac{\mu}{\Lambda}\right)^{2\aoac}
	\right]
        \ \ \
        \mathop{-\!\!\!-\!\!\!-\!\!\!-%
        \!\!\!\!\!\longrightarrow}%
        _{\Lambda \raw \infty}
        \
	\frac{2av}{(4\pi)^2} \frac{\alc}{\alpha}
        \equiv
        \frac{\lambda_{\star}}{y^4_{\star}}
        \ ,
\label{sol:comp:lam:fix}
\ea
where $y^2_{\star}$ and $\lambda_{\star}$ denote
the PR fixed point values.

We should note that
the presence of gauge interaction is essential
for this renormalizability.
This is of course the consequence of the fact that
nontrivial Higgs-Yukawa system can exist
only in the presence of gauge interaction.
Let us demonstrate this more directly in the NJL model
by considering the limit of switching the gauge interaction off.
By using Eq.~(\ref{sol:QCD}),
we take the limit $\alpha_0 \rightarrow +0$
in Eqs.~(\ref{sol:comp:Yukawa}) and (\ref{sol:comp:lam}) and have
\ba
y^2_{\Lambda}(t)
        &{}&\!\!\!\!\!\!
        \mathop{-\!\!\!-\!\!\!-\!\!\!-%
        \!\!\!\!\!\longrightarrow}%
        _{\alpha_0 \raw +0}
        \
        \frac{1}{2a}\frac{(4\pi)^2}{\ln(\Lambda/\mu)}
        \
        \mathop{-\!\!\!-\!\!\!-\!\!\!-%
        \!\!\!\!\!\longrightarrow}%
        _{\Lambda \raw \infty}
        \
        0 \ ,
\label{sol:comp:Yukawa:fix:trivial}
\\
\lambda_{\Lambda}(t)
        &{}&\!\!\!\!\!\!
        \mathop{-\!\!\!-\!\!\!-\!\!\!-%
        \!\!\!\!\!\longrightarrow}%
        _{\alpha_0 \raw +0}
        \
        \frac{\,v\,}{\,a\,}\frac{(4\pi)^2}{\ln(\Lambda/\mu)}
        \
        \mathop{-\!\!\!-\!\!\!-\!\!\!-%
        \!\!\!\!\!\longrightarrow}%
        _{\Lambda \raw \infty}
        \
        0 \ .
\label{sol:comp:lam:fix:trivial}
\ea
We see that
the renormalized couplings identically vanish
in the continuum limit
and that the NJL model is trivial and hence nonrenormalizable
in the absence of gauge interaction.

\msection{Compositeness Condition for Mass Parameter
\label{sec:mass}}
\setcounter{footnote}{0}

As mentioned in the preceding section,
there are some subtleties
in the compositeness condition for the mass parameter.
This condition is important to see how
the four-fermion coupling constant $G$ determines
the phase structure of the equivalent gauge-Higgs-Yukawa system.
It was argued in ref.~\cite{BKMN:CriticalInst} that
the compositeness condition for the mass parameter
is given by\footnote{
The factor $1/2$ appears in LHS of Eq.~(\ref{compcond:mass:MD})
since we are treating the real scalar field $\sigma$.
}
\be
  \frac{1}{2} \frac{\mmd^2(\tcut)}{y^2(\tcut)}
	= G^{-1}
	= \frac{a}{(4\pi)^2}\frac{\Lambda^2}{g_4(\Lambda)} \ ,
\label{compcond:mass:MD}
\ee
where
$\mmd$ is the mass parameter in the mass-dependent (MD) scheme.
The mass parameter $\mmd^2(t)$ renormalized
at $t=\ln(\mu/\mu_0)$ is related
to $m^2(t)$ in the mass-independent $\MS$ scheme
to the leading order in our approximation as\footnote{
The coefficient of $\mu^2$ term
in Eqs.~(\ref{mass:1loop}) and (\ref{RGE:mass:MD})
depends on the detailed schemes in the MD renormalization.
This scheme dependence is intimately related
to the quadratic divergences in the scalar mass
and can be attributed to the discrepancy of the meaning
of the renormalization scales among various schemes.
In ref.~\cite{BKMN:CriticalInst},
such a scheme dependence was treated
by introducing a parameter $\BKMN$,
which takes the value $\BKMN=2$
if we adopt Wilson's cutoff scheme as an MD renormalization.
We have simply put $\BKMN=2$ here
in Eqs.~(\ref{mass:1loop}) and (\ref{RGE:mass:MD})
since then the renormalization point is
directly identifiable with the cutoff.
\addtocounter{footnote}{-1}
}
\be
\mmd^2(t) = m^2(t) + \frac{2a}{(4\pi)^2} y^2(t)\mu^2 \ ,
\label{mass:1loop}
\ee
while other dimensionless coupling constants
remain the same in both schemes.
The RGE's for these mass parameters are, respectively,
given by\footnotemark
\ba
16\pi^2 \frac{d}{dt} m^2(t)
	&=& 2 a\,y^2(t) m^2(t) \ ,
\label{RGE:mass:MI}
\\
16\pi^2 \frac{d}{dt} \mmd^2(t)
	&=& 2 a\,y^2(t)
	\left[ \mmd^2(t) + 2 \mu^2 \right] \ .
\label{RGE:mass:MD}
\ea

Since we are using the $\MS$ renormalization scheme everywhere,
we have to rewrite this compositeness condition
(\ref{compcond:mass:MD}) in terms of
the mass parameter $m^2$ in the $\MS$ scheme.
If we use the relation (\ref{mass:1loop}) literally,
the compositeness condition (\ref{compcond:mass:MD})
is rewritten into
\be
  \frac{1}{2} \frac{m^2(\tcut)}{y^2(\tcut)}
	= \frac{a}{(4\pi)^2} \Lambda^2
	\left[    \frac{1}{g_4(\Lambda)}
		- 1
	\right] \ ,
\label{compcond:mass:MI}
\ee
which was shown in ref.~\cite{KS:MI} to work well,
like the original one (\ref{compcond:mass:MD}),
to the leading order in $1/\Nc$ expansion
in the Higgs-Yukawa system {\it without} gauge interaction.

In the presence of the gauge interaction, however,
Eq.~(\ref{compcond:mass:MI}) is not correct since
the leading-order relation (\ref{mass:1loop})
between the mass parameters in the $\MS$ and MD schemes
is {\it not} consistent with the RGE's
(\ref{RGE:mass:MI}) and (\ref{RGE:mass:MD}).\footnote{
If we neglect the gauge interaction,
the relation (\ref{mass:1loop}) becomes
an exact leading-order relation in the usual $1/\Nc$ expansion.
It is well-known that the quantities calculated
in the leading order of the usual $1/\Nc$ expansion
satisfy RGE at that order,
and hence receive no ``improvement" by RGE.
}
Generally, such approximate relations
like Eq.~(\ref{mass:1loop}) between running parameters
in two different renormalization schemes
are not RG invariant and hence are valid
only in a restricted region of renormalization points.
Since the compositeness condition relates
the low-energy parameters to those at the cutoff $\Lambda$,
the naive use of the leading-order relation (\ref{mass:1loop})
is problematic.

For illustration,
suppose that we calculate the effective potential
in the $\lambda\phi^4$ theory to the leading-log order
using two different renormalization schemes
${\cal R}_1$ and ${\cal R}_2$.
Then we need to know one-loop $\beta$ functions
and tree-level potential forms in both schemes\cite{BKMN:ImpEff}.
The potentials in ${\cal R}_1$ and ${\cal R}_2$,
in particular the coupling constants $\lambda_1(\mu)$
and $\lambda_2(\mu)$, are generally the same at the tree level.
But the tree level relation $\lambda_1(\mu)=\lambda_2(\mu)$,
for instance, is not invariant under one-loop RGE,
and receives loop corrections of the form
\be
  \lambda_1(\mu)
  =
  \lambda_2(\mu)
  + a \frac{\lambda_2^2(\mu)}{(4\pi)^2} \ln \frac{m^2(\mu)}{\mu^2}
  + b \frac{\lambda_2^3(\mu)}{(4\pi)^4}
    \left( \ln \frac{m^2(\mu)}{\mu^2} \right)^2
  + \cdots \ .
\ee
We see from this that the RG non-invariant relation
$\lambda_1(\mu)=\lambda_2(\mu)$ is valid only around
$\mu^2\simeq m^2(\mu)$ where the loop corrections are small.

Our present leading-order relation (\ref{mass:1loop})
is not RG invariant in the presence of gauge interaction
and is analogous to the above tree-level relation
$\lambda_1(\mu)=\lambda_2(\mu)$.
The gauge interaction effects appear from two-loop levels
and give corrections to Eq.~(\ref{mass:1loop}) in the form
\be
  \mmd^2(t)
  =
  \left[
    m^2(t) + \frac{2a}{(4\pi)^2} y^2(t) \mu^2
  \right]
  + \mbox{(const.)}\times y^2(t) \alpha(t) \mu^2 + \cdots \ .
\ee
{}From this, we understand that
the leading-order relation (\ref{mass:1loop})
is most reliable at $\mu=0$ (or $t=-\infty$).
In other words,
{\it the mass parameters in $\MS$ and {\rm MD} schemes should be
identified with each other at} $\mu=0$.
Such identification at $\mu=0$ is all right
for the fixed gauge-coupling case,
but is problematic in the running gauge-coupling case
since the running coupling $\alpha(t)$ in our approximation
diverges at some finite $\mu=\LQCD$ before reaching $\mu=0$.
This implies that
the relation between $\MS$ and MD masses is in fact
a very non-perturbative, dynamical problem
with respect to the gauge interaction.
This could be expected from the beginning
if we recall the following facts:
the $\MS$ mass parameter $m^2(t)$ does not change sign
under the change of the renormalization point $\mu$
since $m^2(t)$ is multiplicatively renormalized.
But the sign of $m^2(t)$ determines
whether the spontaneous breaking of the symmetry
($\sigma\rightarrow-\sigma$) occurs or not,\footnote{
The sign of the MD mass parameter $\mmd^2(t)$, on the other hand,
signals the symmetry breaking of the effective theory valid around the
scale $\mu$.
So, whether the symmetry of the system is {\it eventually} broken or
not is determined by the sign of $\mmd^2(\mu)$ at $\mu=0$.
This also explains why the $\MS$ and MD mass parameters should be
identified with each other at $\mu=0$.
}
and hence is a very dynamical quantity by itself.
Moreover, if the asymptotically free gauge interaction is present,
which becomes strong in the infrared region,
we expect that the symmetry is always spontaneously broken
just like the chiral symmetry in the actual QCD.
If so, the $\MS$ mass $m^2(t)$ should be sensitive
to the presence of gauge interaction,
in particular, to its infrared behavior.

However,
we actually have no idea about
how the `true' running gauge-coupling $\alpha(t)$ behaves
in the region $0\leq\mu\lsim\LQCD$.
Therefore, we are obliged to choose a scale $\mu=\muc$
at which we rely the leading-order relation (\ref{mass:1loop})
to be somewhere around $\LQCD$ but a bit above it:
we choose
\be
  \muc = O(1) \times \LQCD \ .
\label{def:muc}
\ee

With this understanding,
we improve the leading order relation (\ref{mass:1loop})
into a relation between RG invariant quantities.
The RG invariant combination
can easily be found in the $\MS$ scheme:
from the RGE (\ref{RGE:mass:MI}) as well as
the RGE's (\ref{RGE:QCD}) and (\ref{RGE:Yukawa}),
we find that the combination
\be
\run{\cob} \frac{m^2(t)}{y^2(t)}
\label{inv:MI:running}
\ee
is independent of $t=\ln(\mu/\mu_0)$.
The invariant in the MD scheme is a little bit harder to find.
But we can obtain the following equation
from the RGE (\ref{RGE:mass:MD}):
\ba
\frac{d}{dt}
	\left[ \run{\cob} \frac{\mmd^2(t)}{y^2(t)} \right]
&=&	\frac{2a}{(4\pi)^2} 2 \mu^2 \run{\cob} \ .
\label{RGE:noninv:MD}
\ea
Integrating this equation from some arbitrary point $\tc$
to $t$, we find
a $t$-independent combination in the MD scheme:
\be
\run{\cob}
	\left[
	  \frac{\mmd^2(t)}{y^2(t)}
	- \frac{2a}{(4\pi)^2} \frac{\mu^2}{\Omega_0(t;\tc)}
	\right] \ ,
\label{inv:MD:running}
\ee
where $\Omega_0(t;\tc)$ is defined by an integral
\ba
\Omega_0^{-1}(t;\tc)
	&\equiv& e^{-2t} \run{-\cob}
	\int_{\tc}^t 2 dz e^{2z} \eta^{\cob}(z) \ .
\label{def:Omega0}
\ea

Now that we have found the RG invariants both in the $\MS$ and MD
schemes,
we can improve the leading-order relation (\ref{mass:1loop})
to the relation between the RG invariant quantities.
Let us choose $\tc\equiv\ln(\muc/\mu_0)$
to be a scale (\ref{def:muc}) at which the leading-order
relation (\ref{mass:1loop}) is reliable:
\be
\frac{m^2(\tc)}{y^2(\tc)}
	= \frac{\mmd^2(\tc)}{y^2(\tc)}
	- \frac{2a}{(4\pi)^2} \muc^2 \ .
\label{mass:1loop:tc}
\ee
Then rewriting both sides of this equation
by using the RG invariants (\ref{inv:MI:running})
and (\ref{inv:MD:running}),
we have
\be
\frac{m^2(t)}{y^2(t)}
	= \frac{\mmd^2(t)}{y^2(t)}
	- \frac{2a}{(4\pi)^2} \frac{\mu^2}{\Omega(t;\tc)} \ ,
\label{mass:improved:running}
\ee
where $\Omega(t;\tc)$ is defined by
\ba
\Omega^{-1}(t;\tc)
&\equiv&  \Omega_0^{-1}(t;\tc)
	+
	  \left( \frac{\eta(\tc)}{\eta(t)} \right)^{\cob}
	  \frac{\muc^2}{\mu^2}
\label{def:Omega}
\\
&=&	2 e^{-2t} \alpha^{-\cob}(t)
	\left[
	  \int_{\tc}^t dz e^{2z} \alpha^{\cob}(z)
	+ \int_{-\infty}^{\tc} dz e^{2z} \; \alpha^{\cob}(\tc)
	\right] \ .
\label{def:Omega:final}
\ea
An important point is that
while the original relation (\ref{mass:1loop}) does not include
the gauge interaction corrections,
this improved relation (\ref{mass:improved:running})
does contain them in the RHS in the form of $\Omega^{-1}(t;\tc)$.
It may be amusing to note that
Eq.~(\ref{def:Omega:final})
looks like as if we are using as the running gauge
coupling constant a cutoff form:
$\theta(t-\tc)\alpha(t) + \theta(\tc-t)\alpha(\tc) $.

Having the improved relation between the mass parameters
in the $\MS$ and MD schemes,
we can now write down the compositeness condition
which relates the mass parameter of the gauge-Higgs-Yukawa system
to the four-fermion coupling constant of the gauged NJL model.
Since the relation (\ref{mass:improved:running})
holds at any scale $t$,
we can now set $t$ equal to the cutoff $\tcut=\ln(\Lambda/\mu_0)$.
Then, by substituting it into the
compositeness condition (\ref{compcond:mass:MD}),
we obtain
\be
\frac{m^2(\tcut)}{y^2(\tcut)}
	= \frac{2a}{(4\pi)^2} \Lambda^2
	\left[
	  \frac{1}{g_4(\Lambda)}
	- \frac{1}{\Omega(\tcut;\tc)}
	\right] \ .
\label{compcond:mass:MI:improved}
\ee
This is the final form of our compositeness condition
for the mass parameter.

We can now show that the mass parameter $m^2(t)$
at a low energy point $t$ can be made finite
when $\Lambda \rightarrow \infty$ by
the renormalization of the four fermion coupling constant
$g_4(\Lambda)$, as announced in the previous section.
Using Eq.~(\ref{compcond:mass:MI:improved}) and
the RG invariance of the quantity (\ref{inv:MI:running}),
it immediately follows that
\begin{eqnarray}
m^2(t)
&=&
\frac{2a}{(4\pi)^2} \, y^2(t) \,
\left( \frac{\alpha(\tcut)}{\alpha(t)} \right)^{\cob}
\; \Lambda^2
\left[
\frac{1}{g_4(\Lambda)} - \frac{1}{\Omega(\tcut;\tc)}
\right]
\nonumber\\
&=&
\frac{2(c-b)}{4\pi} \, \alpha(t) \,
\left\{
 \left( \frac{\alpha(\tcut)}{\alpha(t)} \right)^{\cob}
\, \Lambda^2 \,
\left[ 1-
\left(\frac{\alpha(\tcut)}{\alpha(t)} \right)^{1-\cob}
\right]^{-1}
\left[
\frac{1}{g_4(\Lambda)} - \frac{1}{\Omega(\tcut;\tc)}
\right]
\right\}
\ ,
\nonumber\\
\label{compcond:mass:MI:improved:2}
\end{eqnarray}
where Eq.~(\ref{sol:comp:Yukawa}) is used in going to the second
expression. This tells us that
we can keep the low energy mass parameter $m^2(t)$ finite and
independent of $\Lambda$. Namely, as the cutoff $\Lambda$ goes to
infinity, the four fermion coupling constant $g_4(\Lambda)$ should
be adjusted such that the quantity in the curly bracket in the
second expression remains $\Lambda$-independent. Therefore the
renormalization of $g_4(\Lambda)$ is performed as
\be
 \left( \frac{\alpha(\tcut)}{\alpha(t)} \right)^{\cob} \,
\left[ 1-
\left(\frac{\alpha(\tcut)}{\alpha(t)} \right)^{1-\cob}
\right]^{-1}
\, \Lambda^2 \,
\left[
\frac{1}{g_4(\Lambda)} - \frac{1}{\Omega(\tcut;\tc)}
\right]
= \mu^2
\left[
\frac{1}{g_{4R}(\mu)} - \frac{1}{g_{4R}^\ast}
\right]
\, ,
\ee
so that we have a renormalized expression
\be
m^2(t)
=
\frac{2(c-b)}{4\pi} \, \alpha(t) \,
\mu^2
\left[
\frac{1}{g_{4R}(\mu)} - \frac{1}{g_{4R}^\ast}
\right] .
\ee
Here $g_{4R}^\ast$ is a finite constant whose value depends on the
definition of $g_{4R}(\mu)$. This completes the renormalizability
proof of the gauged NJL model discussed in the previous section.

Before closing this section,
we consider
the fixed gauge coupling limit
for later convenience.
As we discussed above,
we can take $\tc \raw -\infty$ ($\muc=0$)
without any problems
in this case,
and
$\Omega(t;\tc)$ defined by Eq.~(\ref{def:Omega})
then reduces to a $t$-independent constant:\footnote{
Eq.~(\ref{def:omega}) is valid only when $\alpha<2\alc$.
Otherwise $\Omega^{-1}(t;\tc)$ diverges as $\tc\rightarrow-\infty$,
but such a case is outside the validity of our weak gauge coupling
approximation.}
\be
  \Omega(t;\tc) \,\,\,
  \mathop{-\!\!\!-\!\!\!-\!\!\!-\!\!\!-\!\!\!-\!\!\!-%
    \!\!\!\!\longrightarrow}%
    _{b\rightarrow0,\,\tc\rightarrow-\infty}
  \,\,\, \omega \equiv 1- \frac{1}{2} \frac{\alpha}{\alc} \ .
\label{def:omega}
\ee
By taking the limit $b \raw 0$
in Eq.~(\ref{compcond:mass:MI:improved:2}),
we obtain
\be
m^2(t)
	=
  \frac{2a}{(4\pi)^2} \,
\left( \frac{\Lambda^2}{\mu^2} \right)^{\omega}
  {y_\Lambda^2(t)} \, \mu^2
	\left[
	  \frac{1}{g_4(\Lambda)}
	- \frac{1}{\omega}
	\right] \ ,
\label{compcond:mass:MI:fix:2}
\ee
where $y_\Lambda(t)$ is the Yukawa coupling given in
Eq.~(\ref{sol:comp:Yukawa:fix}).

\msection{Effective Potential and\\
Renormalization of Gauged NJL Model
\label{sec:effpot}}

We have established the renormalizability of
the gauged NJL model with $c>b$
and clarified how the continuum limit
$\Lambda \raw +\infty$ is taken.
In this section,
we calculate the effective potential of
the ``renormalizable" gauged NJL model
and explicitly perform the renormalization.
We then compare the result with the previous one
obtained by Kondo, Tanabashi and Yamawaki (KTY)
for the fixed gauge coupling case
by solving the ladder SD equation\cite{KTY:PTP}.

Our strategy to obtain
the effective potential of the gauged NJL model
is to improve the potential of the gauge-Higgs-Yukawa system
supplemented with the compositeness condition by using the RGE.
We follow the procedure described in detail in ref.~\cite{BKMN:ImpEff}
and adopt the $\MS$ renormalization scheme.
We hereafter restrict ourselves to the fixed gauge-coupling case
since there the comparison with the previous KTY's result is possible
and every calculation can be done explicitly.

Let us start with the effective potential of
the gauge-Higgs-Yukawa system in the leading order
of our approximation:\footnote{
Here and hereafter we omit the vacuum energy term
since it is sub-leading in the $1/\Nc$ expansion.
}
\be
  V_0(g,y,\lambda,\sigma,m^2\,;\mu)
	= \frac{1}{2} m^2 \sigma^2
	+ \frac{1}{4} \lambda \sigma^4
	- \frac{a}{(4\pi)^2} \MF^4
	  \left[ \ln \frac{\MF^2}{\mu^2} - \frac{3}{2} \right] \ .
\label{v0}
\ee
We note that the scalar field $\sigma$ is the renormalized one
with the condition that its kinetic term is normalized to be unity
and we also introduce the fermion mass in the background $\sigma$ by
\be
\MF \equiv y \sigma \ .
\ee

The RG improvement of the effective potential is achieved
by noting the fact that
the effective potential should be RG invariant:
\be
  V\left( g,y,\lambda,\sigma,m^2\,;\mu \right)
= V\left( \wb{g}(t),\wb{y}(t),\wb{\lambda}(t),
	  \wb{\sigma}(t),\wb{m}^2(t)\,;e^t\mu \right) \ ,
\label{sol:RGE:pot}
\ee
where the barred quantities $\wb{g}(t)$, $\wb{y}(t)$, $\cdots$,
are the renormalized parameters at the scale $e^t\mu$;
\be
  \overline{x}(t) = x(e^t\mu) \qquad \qquad \mbox{for}
  \qquad x=g,y,\lambda, m^2,\sigma \  .
\label{def:bar:x}
\ee
[Do not confuse, \eg, $\wb{g}(t)$ with $g(t)$
in the preceding sections which stands for $g(\mu)$.]
The $t$-dependence of these barred parameters are determined by
Eqs.~(\ref{RGE:QCD})--(\ref{RGE:lam}), (\ref{RGE:mass:MI}) and
\be
  16\pi^2 \frac{d}{dt} \wb{\sigma}(t)
	= - a \wb{y}^2(t) \wb{\sigma}(t)
\label{anom.dim}
\ee
with the initial condition $\wb{x}(t\!=\!0)=x(\mu)$.
Then as was shown in ref.~\cite{BKMN:ImpEff},
the RG improved potential $V$ is given as follows:
first evaluate the running parameters
at $t$ determined by
\be
\wbMF (t)=e^t\mu
\label{cond:imp}
\ee
and then, insert them into the leading-order potential $V_0$.
Namely,
\ba
&&V(g,y,\lambda,\sigma,m^2\,;\mu) \nonum
&&=
	\left.
	V_0\left( \wb{g}(t),\wb{y}(t),\wb{\lambda}(t),
		  \wb{\sigma}(t),\wb{m}^2(t)\,;e^t\mu \right)
	\right\vert_{\wbMF (t)=e^t\mu}
\nonumber\\
&&= \left.\left[
	  \frac{1}{2} \wb{m}^2(t) \wb{\sigma}^2(t)
	+ \frac{1}{4} \wb{\lambda}(t) \wb{\sigma}^4(t)
	- \frac{a}{(4\pi)^2} \wbMF^4(t)
		\left(  \ln \frac{\wbMF^2(t)}{e^{2t}\mu^2}
			- \frac{3}{2}
		\right)
	  \right]\right\vert_{\wbMF (t)=e^t\mu} \ .
\label{ep:form2}
\ea
Note that the condition (\ref{cond:imp}) for $t$ is just chosen such
that the logarithmic term drops out.

In order to find the value of $t$ satisfying
the condition (\ref{cond:imp}) for the improvement,
let us examine the RG evolution of the running fermion mass
$\wbMF^2(t) = \wb{y}^2(t)\wb{\sigma}^2(t)$.
{}From the RGE's (\ref{RGE:QCD}), (\ref{RGE:Yukawa})
and (\ref{anom.dim}), we find the combination
$\left(\wb{\alpha}(t)/\wb{\alpha}(0)\right)^{-c/b}\wbMF^2(t)$
RG-invariant.
In the present fixed gauge-coupling case,
this reduces to $e^{(\alpha/\alc)t}\wbMF^2(t)$
and hence we have
\be
e^{(\aoac) t} \wbMF^2(t) = \wbMF^2(t\!=\!0) = \MF^2(\mu) \ .
\label{cond:imp2}
\ee
Substituting the condition (\ref{cond:imp}),
this determines the desired $t$ as
\be
 e^{t} = \left( \frac{\MF(\mu)}{\mu}
	 \right)^{2/(2+\alpha/\alc)}
  = \left( \frac{\MF(\mu)}{\mu}
	 \right)^{1/(2-\omega)} \ ,
\label{cond:t}
\ee
where $\omega$ is defined in Eq.~(\ref{def:omega}).

Substituting the above $t$ into Eq.~(\ref{ep:form2}),
we obtain the RG improved effective potential
of the gauge-Higgs-Yukawa system.
With the compositeness condition imposed,
it becomes the RG invariant effective potential
of the gauged NJL model:
noting that
$\wb{m}^2(t)\wb{\sigma}^2(t)$ is $t$-independent
by Eqs.~(\ref{RGE:mass:MI}) and (\ref{anom.dim}),
we find the quadratic part in $\wb{\sigma}$
in the potential (\ref{ep:form2}) to give
\ba
  V_2
&\equiv&
    \left.\left[
      \frac{1}{2} \wb{m}^2(t) \wb{\sigma}^2(t)
    \right]\right\vert_{\wbMF (t)=e^t\mu}
  =
    \frac{1}{2} m^2(\mu) \sigma^2(\mu)
\nonumber\\
&=&
  \frac{1}{2} \frac{2a}{(4\pi)^2}
  \ycomp^2(\mu)  \, \mu^2 \,
  \left( \frac{\Lambda^2}{\mu^2} \right)^{\omega}
  \left[
    \frac{1}{g_4(\Lambda)}
    - \frac{1}{\omega}
  \right]
  \, \sigma^2(\mu)\ ,
\label{mass part}
\ea
where in the last step
we have used Eq.~(\ref{compcond:mass:MI:fix:2})
resultant from the compositeness condition for the mass parameter.
We next recall Eq.~(\ref{sol:comp:lam:fix}) which resulted
from the compositeness condition for the coupling constants:
it now reads by Eq.~(\ref{def:bar:x})
\be
  \frac{\wblamcomp(t)}{\wbycomp^4(t)}
=
  \frac{2av}{(4\pi)^2} \frac{\alc}{\alpha}
  \left[
    1 - \left(\frac{e^t\mu}{\Lambda}\right)^{2\aoac}
  \right] \ .
\ee
Using this and Eq.~(\ref{cond:t}),
the quartic part in the potential (\ref{ep:form2})
is evaluated as follows:
\ba
  V_4
&\equiv&
  \left.\left[
    \frac{1}{4} \frac{\wb{\lambda}(t)}{\wb{y}^4(t)} \wbMF^4(t)
    + \frac{3a}{2(4\pi)^2} \wbMF^4(t)
  \right]\right\vert_{\wbMF (t)=e^t\mu} \nonum
\nonum
&=&
    \frac{2av}{4(4\pi)^2} \frac{\alc}{\alpha} \mu^4
    \left[
      \left(\frac{\MF^4(\mu)}{\mu^4}\right)^{1/(2-\omega)}
      - \left(\frac{\mu^2}{\Lambda^2}\right)^{\aoac}
	\frac{\MF^4(\mu)}{\mu^4}
    \right]
\nonum
&& \qquad
    +
    \frac{3av}{2(4\pi)^2} \mu^4
    \left(\frac{\MF^4(\mu)}{\mu^4}\right)^{1/(2-\omega)}
    \ ,
\label{V4}
\ea
where $\MF(\mu)=\ycomp(\mu) \sigma(\mu) $.
Putting Eqs.~(\ref{mass part}) and (\ref{V4}) together,
the potential (\ref{ep:form2}) turns out to be
\ba
\frac{(4\pi)^2}{2a} \frac{V}{\mu^4}
&=&
  \frac{1}{2}
  \ycomp^2(\mu)
  \left( \frac{\Lambda^2}{\mu^2} \right)^{\omega}
  \left[
    \frac{1}{g_4(\Lambda)}
    - \frac{1}{g_4^\ast}
  \right]
  \, \frac{\sigma^2(\mu)}{\mu^2}
\nonum
&& \
   {}+
    \frac{v}{4} \frac{\alc}{\alpha}
    \left[
      \left(\frac{\ycomp(\mu) \sigma(\mu) }{\mu}\right)^{4/(2-\omega)}
      -
      \left(\frac{\mu^2}{\Lambda^2}\right)^{\aoac}
      \left(\frac{\ycomp(\mu) \sigma(\mu) }{\mu}\right)^4
    \right]
\nonum
&& \quad
   {}+
    \frac{3}{4}
    \left(\frac{\ycomp(\mu) \sigma(\mu) }{\mu}\right)^{4/(2-\omega)}\ ,
\label{eff:pot}
\ea
where
\be
  g_4^\ast \equiv \omega \ .
\ee
This is the desired RG-improved effective potential
of the gauged NJL model with cutoff $\Lambda$.
This potential $V$ satisfies the RGE
and hence is independent of $\mu$.

We can now see explicitly how
the effective potential is renormalized
when the cutoff $\Lambda$ goes to infinity.
As discussed in section~\ref{sec:mass}, the quadratic term $V_2$
can be renormalized by the four-fermion coupling renormalization:
\be
  \ycomp^2(\mu)
  \left( \frac{\Lambda^2}{\mu^2} \right)^{\omega}
  \left[
    \frac{1}{g_4(\Lambda)}
    - \frac{1}{g_4^\ast}
  \right]
  =
  y^2_\star
  \left[
    \frac{1}{g_{4R}(\mu)}
    - \frac{1}{g_{4R}^\ast}
  \right] \ ,
\label{ren:g4}
\ee
where $y^2_\star$ denotes the PR fixed point value given by
Eq.~(\ref{sol:comp:Yukawa:fix}).
There is still a term containing $\Lambda$
explicitly in Eq.~(\ref{eff:pot}),
the last quartic term proportional
to $(\mu^2/\Lambda^2)^{\alpha/\alc}$,
but it drops out as $\Lambda \raw \infty$ since $\alpha/\alc>0$.
We thus obtain the following finite expression
\be
  \frac{(4\pi)^2}{2a} \frac{V}{\mu^4}
  =
  \frac{1}{2}
  \left(
    \frac{1}{g_{4R}(\mu)}
    - \frac{1}{g_{4R}^\ast}
  \right)
  \frac{y^2_\star\sigma^2(\mu)}{\mu^2}
  + \frac{v}{4} \frac{\alc}{\alpha}
  \left( 1 + \frac{3}{v}\frac{\alpha}{\alc} \right)
  \left( \frac{y_\star \sigma(\mu) }{\mu} \right)^{4/(2-\omega)} \ ,
\label{eff:pot:final}\label{eff:ren-pot}
\ee
giving the final form of
the renormalized effective potential in the gauged NJL model.

Some remarks are in order here.
As we have discussed in section \ref{sec:renorm},
the presence of gauge interaction is essential for
the interaction term to exist nontrivially
as in Eq.~(\ref{eff:ren-pot}).
Actually,
if the gauge interaction is absent,
we should take $\alpha \raw 0$ in Eq.~(\ref{eff:pot})
{\it before} taking $\Lambda \raw \infty$ limit and find
\ba
V
&=&
  \frac{1}{2} \,
  \frac{1}{\ln(\Lambda/\mu)}
  \Lambda^2
  \left[
    \frac{1}{g_4(\Lambda)}
    - \frac{1}{g_4^\ast}
  \right]
  \, \sigma^2(\mu)
\nonum
&&{}+
    \frac{v}{4}
    \frac{(4\pi)^2}{2a}
    \left( \frac{1}{\ln(\Lambda/\mu)} \right)^2
    \left[
      2 \ln(\Lambda/\mu)
      -\ln \left(
         \frac{(4\pi)^2}{2a} \frac{\sigma^2(\mu)}{\mu^2\ln(\Lambda/\mu)}
      \right)
    \right]
    \sigma^4(\mu)
\nonum
&& \
   {}+
    \frac{3}{4}
    \frac{(4\pi)^2}{2a}
    \left( \frac{1}{\ln(\Lambda/\mu)} \right)^2
    \sigma^4(\mu) \, .
\label{eff:pot:nogaugelimit}
\ea
The quadratic term can be made finite
again by renormalizing $g_4(\Lambda)$ suitably, but
the quartic term necessarily vanishes as $\Lambda \rightarrow \infty$,
and this clearly shows the triviality of the pure NJL model.
[The Yukawa term also vanishes as $(1/\ln(\Lambda/\mu))^{1/2}$.
See Eq.~(\ref{sol:comp:Yukawa:fix:trivial}).]

It may be interesting to see how this triviality is related with the
usual knowledge of non-renormalizability of the NJL model.
Recall that we have used the field variable $\sigma(\mu)$ which has a
normalized kinetic term.
If we had used the variable
$\widetilde{\sigma}^2(\mu) \equiv \sigma^2(\mu)/\ln(\Lambda/\mu)
\propto \MF^2(\mu)$ and regarded it as finite ($\Lambda$-independent),
then, we would obtain a finite Yukawa term,
but would encounter logarithmic divergences in the scalar kinetic and
quartic terms.

Our effective potential (\ref{eff:pot})
and its renormalized version (\ref{eff:ren-pot})
should be compared with the previous results
by Bardeen and Love\cite{Bardeen-Love} and by KTY\cite{KTY:PTP},
respectively,
which were obtained by quite a different method
using the SD equation in the ladder approximation.
Our result takes precisely the same form as theirs
under a suitable translation of notations:
in particular our $\omega$, Eq.~(\ref{def:omega}),
is identified with KTY's $\omega=\sqrt{1-\alpha/\alc}$.
Both $\omega$ are of course the same in the first nontrivial order
in the gauge coupling expansion
in which we are working.

The potential (\ref{eff:pot:final}) tells us the following:
it is the sign of the coefficient of the quadratic term
that determines whether
the spontaneous breaking of the symmetry $(\sigma\raw-\sigma)$
occurs or not.
By Eq.~(\ref{ren:g4}),
it is the same as the sign of
$\left[ 1/g_4(\Lambda) - 1/g_4^\ast \right]$.
Therefore the critical value for
the original four-fermion coupling constant $g_4(\Lambda)$
is given by $g_4^\ast=\omega$, Eq.~(\ref{def:omega}).
The critical coupling given by KTY was $(1+\omega)^2/4$
which also coincides with our result $\omega$
in the order $\alpha$ of the present approximation.

Finally, we can do the same calculations
for the running gauge-coupling case also.
There appears no essential difficulties,
but the expression necessarily
becomes implicit there since the value $t$ determined by
Eq.~(\ref{cond:imp}) cannot be solved explicitly; it leads to, in
place of (\ref{cond:imp2}),
\be
 \left( 1 + \frac{b\alpha_0}{2\pi} t \right)^{\cob} e^{2t}
 =  {\MF^2(\mu)\over \mu^2} \ .
\ee
We here comment only on the critical four-fermion coupling constant,
which can be given explicitly even in this running gauge-coupling case.
The symmetry breaking is judged by the sign of the quadratic term $V_2$
in the effective potential $V$, which now reads,
by using Eq.~(\ref{compcond:mass:MI:improved:2}),
\ba
  \!\!\!\!\!\!\!\!\!\!\!
  \frac{1}{2} \wb{m}^2(t) \wb{\sigma}^2(t)
&=&
 \frac{1}{2} \frac{m^2(\mu)}{y^2(\mu)} \MF^2(\mu) \ .
\nonumber\\
&=&
  \frac{1}{2} \frac{2a}{(4\pi)^2} \Lambda^2
  \left(
    1 + \frac{b\alpha(\mu)}{2\pi} \ln\frac{\Lambda}{\mu}
  \right)^{-\cob}
  \left[
    \frac{1}{g_4(\Lambda)}
    - \frac{1}{\Omega(\tcut;\tc)}
  \right]
  \MF^2(\mu) \ .
\label{mass part2}
\ea
Therefore the critical coupling for $g_4(\Lambda)$
is given in this case  by
\be
g_4^* = \Omega(\tcut;\tc)\ .
\ee
If the cutoff is large enough such that
$\Lambda \gg \muc \simeq \LQCD$ and $\alpha(\Lambda) \ll 1$, \
this $\Omega(\tcut;\tc)$ is evaluated by Eq.~(\ref{def:Omega:final})
to be
\be
\Omega^{-1}(\tcut;\tc) = 1 + \frac{\alpha(\Lambda)}{2\alc}
+ O\Big( \alpha^2(\Lambda), \ \frac{\muc^2}{\Lambda^2} \Big) \ .
\ee
This again agrees with the previous result reported in
Refs.~\cite{BKS,KSY,BKMN:CriticalInst}.

\msection{Conclusions}

We have explored in this paper
the nontriviality constraint for the gauge-Higgs-Yukawa system
and revealed the low-energy structure of
the theory.
The requirements of nontriviality and stability constrain the low
energy Yukawa and quartic scalar couplings to lie on a line connecting
``Gaussian'' and the PR fixed points.
The upper bound (the PR fixed point) can be apart from the
``Gaussian'' point only in the presence of gauge interaction.

The gauged NJL model is equivalent with the gauge-Higgs-Yukawa system
supplemented by the compositeness condition.
The equivalent gauge-Higgs-Yukawa theory with cutoff $\Lambda$
approaches the PR fixed point as $\Lambda\rightarrow\infty$ from
outside the allowed region.
Thus the continuum gauged NJL model lies on the boundary of the
allowed region of the nontrivial gauge-Higgs-Yukawa system.
This proves the renormalizability of gauged NJL model.

We have also calculated the effective potential of the gauged NJL
model by using RGE.
It is interesting that the result agrees with the KTY's one which was
obtained by quite a different method, the ladder SD equation.

In the model considered here, the scalar field was gauge-singlet.
It may be interesting to investigate the case in which
the gauge group is extended to a semi-simple one just like
$SU(3)_{\rm C} \times SU(2)_{\rm L}$ in the SM,
so that the scalar field becomes gauge-non-singlet.
We expect that our main conclusions concerning the nontriviality
and the renormalizability will remain true,
but the coupling reduction of $\lambda$ to $y$ might no longer occur.

\paragraph{Note Added:} We are informed that Gusynin and Miransky
\cite{G-M} have derived a renormalized
low-energy effective action in the broken phase of the gauged-NJL model
(with fixed gauge coupling). See also the discussion
for the symmetric phase at $\alpha\simeq\alpha_{\rm c}$
given in Miransky\cite{Miransky}.

\section*{Acknowledgement}

The authors would like to express their sincere thanks to K.~Yamawaki
who gave an excellent series of lectures at Kyoto University and
suggested this problem to them.
They are also indebted to M.~Bando and T.~Maskawa
for discussions and comments.
M.~H.\ and T.~K.\ are
supported in part by the Grants-in-Aid for Scientific
Research
\#2208 and \#06640387, respectively,
from the Ministry of Education, Science and Culture.

\ifnum\figcond>0
  \ifnum \figsty>0 \figepsfout
\fi\fi

\end{document}